\documentclass[fleqn,usenatbib]{mnras}

\usepackage[T1]{fontenc}
\usepackage{ae,aecompl}
\usepackage{hyperref}
\usepackage[toc,page]{appendix}

\usepackage{graphicx}	
\graphicspath{{images/}}
\usepackage{amsmath}	
\usepackage{amssymb}	
\usepackage{cancel} 
\usepackage{booktabs,multirow} 
\usepackage{soul}
\usepackage{xcolor}
\usepackage{pdfpages}

\usepackage{newtxtext,newtxmath}

\definecolor{cadmiumred}{rgb}{0.89, 0.0, 0.13}

\definecolor{burgundy}{rgb}{0.5, 0.0, 0.13}

\definecolor{scc}{rgb}{0.13, 0.67,  0.8}

\definecolor{what}{rgb}{0.7, 0.2, 0.9}

\setcitestyle{comma}

\title[Reliable photometric redshifts using ML]{Improving the reliability of photometric redshift with machine learning}

\author[O. Razim et al.]{Oleksandra Razim$^{1}$\thanks{E-mail: shr.razim@gmail.com},
Stefano Cavuoti$^{2,1}$,
Massimo Brescia$^{2}$, \and
Giuseppe Riccio$^{2}$,
Mara Salvato$^{3}$,
Giuseppe Longo$^{1}$
\\
$^{1}$Department of Physics, University Federico II, Strada Vicinale Cupa Cintia, 21, 80126 Napoli, Italy\\
$^{2}$INAF - Astronomical Observatory of Capodimonte, Salita Moiariello 16, I-80131 Napoli, Italy\\
$^{3}$MPI for Extraterrestrial Physics, Giessenbachstrasse 1, Garching b. M\'uchen, 85748, Germany
}

\date{Accepted XXX. Received YYY; in original form ZZZ}

\pubyear{2021}

\begin{document}
\label{firstpage}
\pagerange{\pageref{firstpage}--\pageref{lastpage}}
\maketitle

\begin{abstract}
In order to answer the open questions of modern cosmology and galaxy evolution theory, robust algorithms for calculating photometric redshifts (photo-z) for very large samples of galaxies are needed. Correct estimation of the various \mbox{photo-z} algorithms' performance requires attention to both the performance metrics and the data used for the estimation. In this work, we use the supervised machine learning algorithm MLPQNA to calculate photometric redshifts for the galaxies in the COSMOS2015 catalogue and the unsupervised Self-Organizing Maps (SOM) to determine the reliability of the resulting estimates. We find that for $z_{\mathrm{spec}}<1.2$, MLPQNA \mbox{photo-z} predictions are on the same level of quality as SED fitting photo-z. We show that the SOM successfully detects unreliable $z_{\mathrm{spec}}$ that cause biases in the estimation of the \mbox{photo-z} algorithms' performance. Additionally, we use SOM to select the objects with reliable \mbox{photo-z} predictions. Our cleaning procedures allow to extract the subset of objects for which the quality of the final \mbox{photo-z} catalogs is improved by a factor of two, compared to the overall statistics.

\end{abstract}



\section{Introduction}

Knowledge of spatial distribution of galaxies is crucial for answering most of the open questions in observational cosmology and galaxy evolution. In particular, we need to know galaxy distances. The universal method of deriving them is to measure redshifts of galaxies, caused by the cosmological expansion. The most accurate way of doing that is to calculate spectroscopic redshifts (spec-z, or $z_{\mathrm{spec}}$), but this method requires high-quality spectra for all investigated objects. Obtaining these spectra is time-consuming and sometimes impossible for the faint sources.
Currently, wide-field galaxy surveys cannot provide \mbox{spec-z} catalogues of the same depth and richness as photometric catalogues. For example, for the 16-th Data Release of the Sloan Digital Sky Survey (SDSS) the ratio between the number of detected galaxies and the number of available spectroscopic redshifts is of the order of $1/100$\footnote{Based on the sizes of photometric and spectroscopic catalogues obtained via \url{http://skyserver.sdss.org/CasJobs/}}, going down to even smaller numbers for other surveys. 

This difficulty of deriving galaxy redshifts from spectra led to the development of alternative techniques, collectively called \textit{photometric redshift methods} (photo-z, or $z_{\mathrm{phot}}$), first proposed in \citealt{baum1957,baum1962} and further investigated in a vast number of publications, e.g. \citealt{baldwin1977,Butchins1980,Koo1985,connolly1995,Gwyn1996,Bolzonella2000,Benitez2000,Collister2004,Ball2008,Brammer2008,Gerdes2010,Carrasco2013,Carrasco2014,Hoyle2016,Bonnett2016,Sadeh2016,Bilicky2018,Pasquet2019} and many others.

These methods are based on the fact that multi-band photometry can be treated as a low-resolution spectrum and hence as an approximation of the intrinsic Spectral Energy Distribution (SED). 
Cosmological redshift causes a stretching of the SED and a change in the wavelengths of the prominent spectral features (Lyman or Balmer breaks, emission and absorption lines, etc.) that, moving in or out of the fixed bands of a given photometric system, cause changes in magnitudes and colours. From these changes we can derive \mbox{photo-z}.

There is a variety of \mbox{photo-z} estimation methods. The majority of them uses one of the two main approaches: a theoretical one, called SED template fitting, and an empirical one, largely based on machine learning (ML) paradigms \citep{Desprez,Schmidt2020}.

SED template fitting methods use spectral template libraries obtained from either observations or galaxy models. These templates are shifted across the expected redshift range (using an arbitrarily chosen step) and then convoluted with the transmission curves of the acquisition system in order to derive estimated magnitudes. The resulting magnitudes are then fitted to the observed data in order to find the combination of a template and redshift value which minimise the residuals. 
Depending on the particular approach, template fitting methods may include such steps as producing new templates via interpolation between the already existing ones, correcting systematic biases of the observed photometry, compensating for the absorption of the interstellar medium, etc.
A downside of these methods is that they require template libraries that are conditioned by our knowledge of galaxy evolution and therefore by definition incomplete.

Instead of using pre-defined templates, ML derives the correlation between photometry and redshift from the data themselves. To train the ML models we need a knowledge base (KB), formed by both photometry and high-quality \mbox{spec-z} for a considerable number of objects (how large depends on the homogeneity of the sample under scrutiny). With a KB large enough and fully covering the  parameter space, the ML approach allows to take into account all observational and physical effects automatically \citep{Brescia2018}. 
At the same time, providing a high-quality KB is not a trivial task by itself. For instance, due to the fact that ML methods have a limited extrapolating power, they cannot provide reliable \mbox{photo-z} predictions outside the spectroscopic range covered by the KB. This is particularly relevant in the faint object domain, usually not properly covered by spectroscopic surveys.

To provide best performance, SED fitting and ML require different types of data, and for this reason the areas of application of the two methods are slightly different \citep{Salvato2019}. In the low-redshift regime \mbox{ML photo-z} often perform better than SED fitting methods, provided that a well-representative spectroscopic KB exists. For high-redshift objects, SED template fitting usually shows better results, mostly due to the lack of a consistent KB for these galaxies. Furthermore, in the case of severe depth imbalance between spectroscopic and photometric information, SED fitting can be the only option. 

Lately, hybrid \mbox{photo-z} techniques that unite strengths of both approaches have appeared (see e.g. \citealt{Newman2008, Beck2016, Cavuoti2017, Duncan2018}). Still, at this point no method allows to estimate \mbox{photo-z} with the same precision as high quality spec-z. Currently, the accuracy of spectroscopic redshift surveys can reach an error of $\sigma \sim 10^{-3}$, where $\sigma$ is a standard deviation of the distribution of residuals between repeated measurements of \mbox{spec-z} of the same objects (e.g. \citealt{LeFevre2005,Hasinger2018}).
The accuracy of \mbox{photo-z} catalogues obtained with broad-band photometry is characterized by $\sigma \sim 0.02$ at best \citep{Euclid,LSST,Salvato2019,Brescia2014}, where $\sigma$ describes the distribution of residuals between \mbox{spec-z} and photo-z. Nevertheless, for the upcoming massive surveys, such as the Rubin Observatory Legacy Survey of Space and Time (LSST, \citealt{LSST2019}) and the Euclid (\citealt{Euclid}), deriving \mbox{photo-z} is the only realistic way to obtain distances for the majority of the observed galaxies, so further improvements of \mbox{photo-z} methods are required.

Among the many factors that affect the quality of the \mbox{photo-z} are the issues related to the completeness and quality of the spectroscopic catalogues used for SED fitting calibration and ML training. Firstly, incompleteness of \mbox{spec-z} sample limits the performance of \mbox{photo-z} algorithms, especially the ML ones. Usually \mbox{spec-z} catalogues are incomplete in the faint part of the magnitude parameter space, but the selection function of the spec-z observations may affect the performance in the whole range of magnitudes. Second, the miscalculated \mbox{spec-z} affect the reliability of the performance metrics used for the estimation of quality of photo-z. Some percentage of miscalculated or misidentified sources is present even in the \mbox{spec-z} samples that are believed to be reliable. Typical high-confidence redshift quality flags provide us with 95-99\% of reliable sources, implying that 1-5\% of sources have unreliable spec-z. Consequently, there is no way to say which of the \mbox{photo-z} outliers are due to miscalculated photo-z and which are related to incorrect \mbox{spec-z} values.

For small-size surveys the objects with noticeable difference between \mbox{spec-z} and \mbox{photo-z} can be manually inspected (see e.g. \citealt{Lanzetta1998} and \citealt{Masters2017}), but for large-scale surveys this is not feasible. As a result, we have to find a way of disentangling the various contributions to the error budget, i.e. to distinguish the \mbox{photo-z} prediction error induced by the defects of an estimation method from the uncertainty carried by the contamination of the spectroscopic data.

In this work we present a data cleaning methodology, focused on handling these two data-related issues, namely:
\begin{itemize}
    \item identifying unreliable spec-z;
    \item deriving from the photometric catalogue a set of objects that can be considered as well-represented by the \mbox{spec-z} catalogue, and therefore trusted in terms of \mbox{photo-z} quality estimations.
\end{itemize}

As we will show, both procedures allow to significantly improve the quality of the final \mbox{photo-z} catalogues. 

Our methodology is based on the Self-Organizing Map (SOM) algorithm, first proposed in \cite{Kohonen1982}, and on an approach proposed in \citet{Masters2015,Masters2017}. For the \mbox{photo-z} calculation we use 30-band photometric data from the COSMOS2015 catalogue \citep{Laigle2016}, the master \mbox{spec-z} catalogue provided by COSMOS collaboration (obtained via private communication with Mara Salvato), and the Deep Imaging Multi-Object Spectrograph (DEIMOS) \mbox{spec-z} catalogue \citep{Hasinger2018}. To obtain \mbox{photo-z} we apply the Multi Layer Perceptron with Quasi Newton Algorithm (MLPQNA, \citealt{Brescia2013,Brescia2014}). For additional testing of our data cleaning methodology we also use the SED template fitting \mbox{photo-z} catalogue described in \cite{Laigle2016}. 

This paper is organized as follows. Section~\ref{baseCleanData} explains how we obtain and pre-process our data. 
In Section~\ref{methods} we describe our algorithms, and in Section~\ref{experiments} we describe the experiments on calculating the \mbox{photo-z} and improving their quality with the SOM. Section~\ref{discussion} contains the discussion of the results. The conclusions are stated in section~\ref{conclusions}. The Jupyter notebooks necessary to reproduce the work are available for download via GitHub repository \url{https://github.com/ShrRa/COSMOS_SOM}. MLPQNA is available as a part of PhotoRAPTOR software \citep{photoraptor}. The final \mbox{photo-z} catalogue will be published via CDS Vizier facility.
\section{Data} \label{baseCleanData}
In this section we describe the three catalogues used in this work:
\begin{enumerate}
    \item COSMOS2015, which is our source of photometric measurements and SED fitting photo-z. Its detailed description can be found in \cite{Laigle2016}.
    \item A compilation of spectroscopic redshifts available in literature for the same COSMOS field. We will call this catalog our main spec-z catalog.
    \item The Deep Imaging Multi-Object Spectrograph (DEIMOS) \mbox{spec-z} catalogue \citep{Hasinger2018}, used to perform an additional, independent test of our methods. Same as the main \mbox{spec-z} catalogue, the DEIMOS sample covers most of the COSMOS2015 field.
\end{enumerate}

\subsection{COSMOS2015 photometric catalogue}

The COSMOS2015 catalogue contains multi-wavelength broad-range (from mid-IR to near-UV) photometry for about half million objects, for which X-Ray and radio measurements, star formation rates and other additional information is available. Moreover, this catalogue includes SED fitting \mbox{photo-z}s, obtained with \texttt{LePHARE} software \citep{Arnouts1999,Ilbert2006}.

One of the most important aspects to consider in any ML based experiment is the selection of the training parameter space, i.e. input features that maximize the amount of suitable information. In the case of ML methods for \mbox{photo-z} this problem has been widely discussed  (cf. \citealt{DIsanto2018,Brescia2018}, and references therein). From the ML point of view, an automatic feature selection is not a trivial task: many algorithms for feature selection exist and have been tested on the problem (see \citealt{Donalek2013,Baron2019,Brescia2019}), as well as some data driven brute force approaches (\citealt{Polsterer,DIsanto2018}), which may provide optimal results at the price of very high computational costs, making it impossible to use it on massive and high-dimensional datasets.

In this work, we do not perform an all encompassing feature selection investigation. Instead, we try several feature sets, based on general physical considerations and our past experience. To try these different feature sets, we downloaded the COSMOS2015 photometric data consisting of $34$ bands:
\begin{itemize}
    \item UV broad and medium bands: \texttt{NUVmag, FUVmag, u};
    \item optical and near-IR broad bands: \texttt{B, V, ip, r, zp, zpp, Ks, Y, H, J, Hw, Ksw, yH};
    \item mid-IR broad bands: \texttt{3\_6mag, 4\_5mag, 5\_8mag, 8\_0mag};
    \item optical and near-IR medium bands: \texttt{IA484, IA527, IA624, IA679, IA738, IA767, IB427, IB464, IB505, IB574, IB709, IB827};
    \item and optical and near-IR narrow bands: \texttt{NB711, NB816}.
\end{itemize}

For all of them $2''$ and $3''$ apertures were available, except for Spitzer
Large Area Survey with Hyper-Suprime-Cam (SPLASH; \texttt{3\_6mag, 4\_5mag, 5\_8mag, 8\_0mag}) and GALEX \texttt{(NUVmag, FUVmag)} bands. 

The pre-processing of this dataset consisted of the following steps:

\begin{enumerate}
    \item  To ensure homogeneity of photometry and to enable a comparison with SED fitting photo-z, we follow the procedure described in \citealt{Laigle2016}. We consider only those objects that lay within both UltraVISTA (this is done by using the condition \texttt{Area==0}) and COSMOS (\texttt{Cfl==1}) sky areas (see \citealt{McCracken2012},  \citealt{Capak2007} and \citealt{Scoville2007} for a detailed descriptions of these regions). 
    As a result we extract $576,762$ objects out of the initial $1,182,108$ samples available.
    \item We also exclude stars, X-ray and unclassified sources (\texttt{OType==0}). This leaves us with $551,538$ objects.
    \item Finally, we remove saturated sources, by rejecting objects masked in optical broad bands (\texttt{Sat==0}). After this step, the final photometric catalogue consists of $518,404$ objects.
\end{enumerate}


\subsection{Main Spectroscopic catalogue} \label{spec-z_prep}

The main spec-z catalog is extracted from the spectroscopic COSMOS master catalog maintained within the COSMOS collaboration and it includes only the publicly available redshifts prior Fall 2017.
This catalogue contains $65,426$  spectral redshifts, obtained with 27 different instruments in a \mbox{spec-z} range $0<z_{\mathrm{spec}}<6.5$. The pre-processing consisted of the following steps:

\begin{enumerate}
    \item To exclude stars, we remove sources with $z_{\mathrm{spec}}<0.01$; we also remove objects with $z_{\mathrm{spec}}>9$ to discard likely erroneous spec-z;
    \item AGNs often pose a contamination problem for \mbox{photo-z} algorithms \citep{2019PASP..131j8004N}. Therefore, we remove sources visible in X-Ray, using a catalogue of AGN sources detected by Chandra in the COSMOS field \citep{Civano2012}.
    \item Then we clean the resulting \mbox{spec-z} catalogue from unreliable instances, using the available quality flags \texttt{Q\_f}, described in \cite{Lilly2009}. We select only robust spectroscopic redshifts (i.e. with $\sim 99.6$\% of spectroscopic verification), using the conditions \texttt{2<Q\_f<5} and \texttt{22<Q\_f<25}. 
\end{enumerate}

It is important to note that the Main \mbox{spec-z} catalogue is a compilation of multiple catalogues. These data were obtained during different surveys with different targeting strategies and quality requirements during the last two decades, so the exact quality of the spectroscopic verification is impossible to estimate. As a result, the actual robustness of the final \mbox{spec-z} set may be lower, and this is one of the issues that we will address in the following sections. Another nuance is that for some objects the main \mbox{spec-z} catalogue contains multiple measurements made with different instruments, and in some cases these \mbox{spec-z} values have large residuals between each other $(>0.1)$. At this stage we do try to determine which measurements are correct, neither discard these objects. Instead, during the crossmatch, we simply use the spec-z measurements that are the best coordinate match to the COSMOS2015 objects. In \S~\ref{expBeforeSOM} we analyse these objects to clarify the nature of the \mbox{photo-z} outliers.

The resulting dataset is cross-matched with the COSMOS2015 photometric catalogue, obtaining $\sim20,000$ objects (the exact amount depends on the bands involved for limiting photometric errors, varying in the various experiments; see \S~\ref{Final_prep}). The pre-processing does not noticeably affect the \mbox{spec-z} distribution for $z_{\mathrm{spec}} \leq4$. However, the number of objects for $z_{\mathrm{spec}}>1.2$ is approximately one order of magnitude lesser than the amount of closer objects (see Fig.~\ref{fig:spectrZ-Qf} in Appendix~\ref{AppendA} for the redshift distributions before and after the cleaning). In absolute numbers, we have only $\sim700$ galaxies in the redshift range $1.2<z_{\mathrm{spec}}<4$. Such number of objects is not enough to effectively train our \mbox{photo-z} algorithm in this redshift range, and for this reason we limit our further analysis and the resulting catalogue to $z_{\mathrm{spec}} \leq 1.2$.

\subsection{DEIMOS \mbox{spec-z} catalogue}\label{Deimos_prep}

\begin{figure}
 \includegraphics[width=0.47\textwidth]{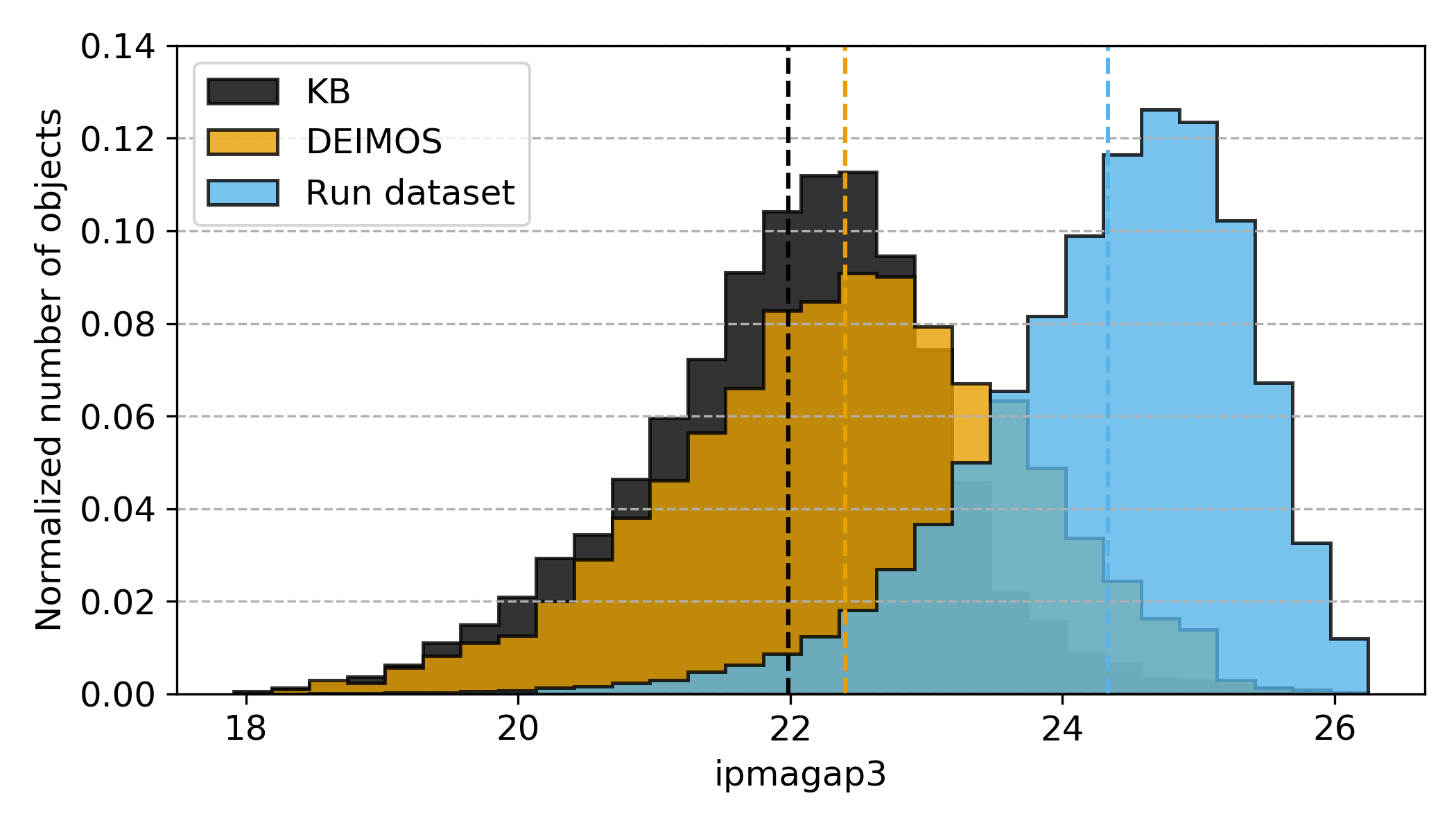}
 \caption{Normalized magnitude distributions for the KB, DEIMOS and run datasets in the \texttt{ipmagap3} after standard cleaning but before any SOM filtering. Dashed lines show the mean value of the distribution.}
 \label{fig:magDistribs}
\end{figure}

Since we want to test our methodology for selecting the subset of photometric data that is well covered by the KB, we need an independent \mbox{spec-z} catalogue different from the main \mbox{spec-z} catalogue selection function. 
For this purpose, we used the catalogue of spectroscopic redshift presented in  \citep{Hasinger2018}, acquired with DEIMOS, within different programs. This catalog provides redshifts for sources that are not included in the main \mbox{spec-z} catalog and that are somewhat fainter (see \mbox{Fig. \ref{fig:magDistribs}}). For these differences, it represents an excellent benchmark data for this study.

In the preparation of the DEIMOS \mbox{spec-z} catalogue, we follow the same procedure described in \S~\ref{spec-z_prep}, aimed at discarding stars, AGNs and unreliable sources\footnote{Note that the DEIMOS catalogue has two different quality flag columns. The one following the same scheme as in \citealt{Lilly2009} is labeled "Qf".}.

\subsection{Final catalogues and SOM data cleaning prerequisites}\label{Final_prep}

\begin{figure}
\centering
 \includegraphics[width=0.4\textwidth]{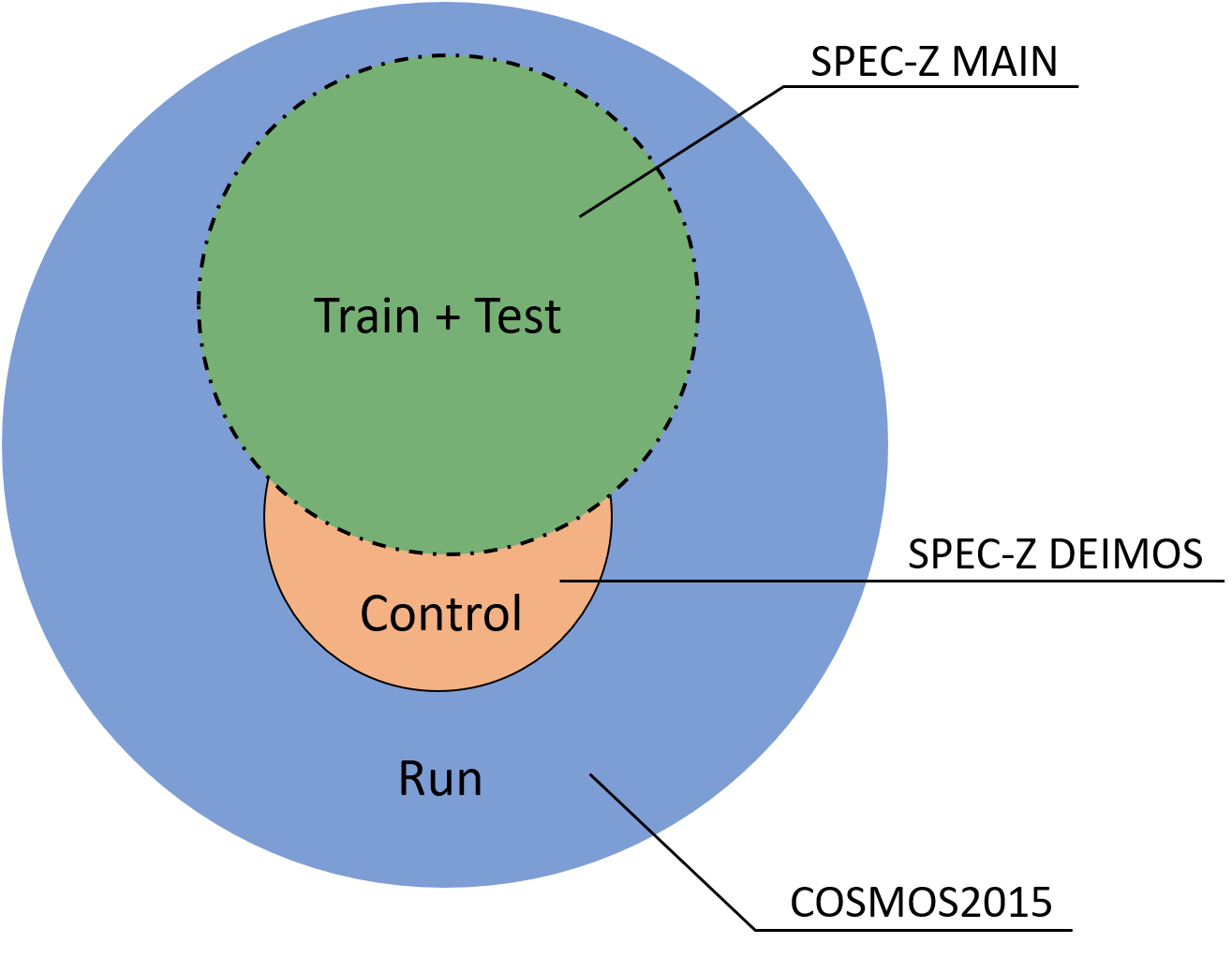}
 \caption{Venn diagram of the catalogues used in this work.}
 \label{fig:catalogue_diagram}
\end{figure}

After the basic preprocessing described in the previous sections, we produce the following datasets (see \mbox{Fig. \ref{fig:catalogue_diagram}}):

\begin{enumerate}
    \item KB (knowledge base), which is the intersection crossmatch between the COSMOS2015 and main \mbox{spec-z} catalogues. It contains both photometry and spec-z. For the \mbox{ML photo-z} experiments we randomly split this KB into train (70\% of KB) and blind test (30\%) datasets to provide reliable evaluation of the model performance (see Fig.~\ref{fig:learningCurve} in Appendix~\ref{AppendA} for the comparison of the model performance on the train and test datasets). To avoid overfitting we also use 10-fold cross validation during the training;
    \item Run dataset, which is the COSMOS2015 catalogue after excluding the objects from the KB. It contains only photometric data;
    \item DEIMOS dataset, which is an intersection of the run dataset and DEIMOS \mbox{spec-z} catalogue (meaning that it does not contain the objects from the KB). We use the DEIMOS as a control dataset to check how well our cleaning procedures work on independent data, occupying a different hypervolume in the parameter space (see \S~\ref{Deimos_prep}).
\end{enumerate}

\begin{table*}
 \begin{tabular}[width=0.8\textwidth]{c c c c c c} 
 \hline
 Dataset & Size & spec-z range & Median spec-z & \texttt{imagap3} range & Median \texttt{imagap3} \\[0.5ex] 
 \hline\hline
 KB  & 19893 & [0.020; 1.2]& 0.57 & [17.91; 26.19] & 22.02 \\ \hline
 DEIMOS  & 2288 & [0.025; 5.68] & 0.89 & [18.48; 25.90] & 22.93 \\ \hline
 Run  & 228361 & NA & NA & [18.14; 26.25] & 24.49\\ \hline
 \end{tabular}
     \caption{Main spectroscopic properties of the datasets. For the initial set of experiments, described in \S~\ref{expBeforeSOM}, different sets of bands were used. The filtering of the objects with large magnitude errors was performed for each feature set independently (see~\S~\ref{Final_prep}), meaning that the exact parameters of the datasets were varying from experiment to experiment. However, these variations were small (e.g. the variations of the sizes of the datasets were less than 1\%). The parameters given in this table describe the dataset with an optimal feature set, which was determined in \S~\ref{expBeforeSOM} and then used throughout the rest of the paper.}
 \label{table:datasets_params}
\end{table*}

In order to ensure the quality of the trained model, a reliable photometry is required, so we excluded all objects with high magnitude errors. Some bands have too many objects with unreliable photometry; if applied to these bands, this part of the pre-processing would reduce the size of the dataset by a factor of two or more, and narrow the area of parameter space where the \mbox{photo-z} algorithm would be applicable. To avoid this, we identified the bands affected by too many objects with large magnitude errors ($e\_mag>1$) and excluded them from the experiments. As a rule of thumb this selection was operated by removing bands containing one order higher number of unreliable measurements than the others; these bands are \texttt{5\_8mag, 8\_0mag, NUVmag and FUVmag} (see Fig.~\ref{fig:magErr} in Appendix~\ref{AppendA}). They contain thousands of objects with large magnitude uncertainties, while the other bands have up to hundreds of such galaxies.

Afterwards, we clean the photometry in the remaining bands, by limiting the magnitude error within the range $0<e\_mag<1$, where $e\_mag$ are magnitude errors for each band. The cleaning is performed for every feature set separately, since it allows us to preserve as many objects as possible (for example, in the experiments where we use only broad bands we do not remove the objects with high magnitude errors in narrow bands). Then in all ML experiments we independently normalize each band to the $[0,1]$ range.

In order to take care of the mentioned low extrapolative power of ML models, we have to make sure that the run dataset is compliant with the KB in terms of parameter space coverage.
\mbox{Fig. \ref{fig:magDistribs}} shows the magnitude distribution for the KB, DEIMOS and run datasets in \texttt{ipmagap3}. As it can be seen, our run dataset is noticeably deeper than the KB and DEIMOS datasets, as expected due to the spectroscopic bias induced by the targeting algorithms.

We then limit the magnitudes in the run dataset bands by their corresponding maximum values present within the KB. However, this is just a preliminary solution, since it does not guarantee a full similarity between the KB and run datasets, with respect to the parameter space. In \S~\ref{methodSOMphotometry} we introduce a more accurate procedure to align the run dataset to the parameter space of the KB.

The summary on the parameters of the datasets used in this work (except the preliminary experiments to determine the optimal feature set, described in \S~\ref{expBeforeSOM}) is given in Tab.~\ref{table:datasets_params}.

\section{Methods} \label{methods}

In this work, we use two neural network algorithms. 
To calculate the \mbox{photo-z} point estimations, we use the supervised ML algorithm MLPQNA \citep{Brescia2013,Brescia2014}. Furthermore, in order to investigate and clean the datasets, we use an unsupervised ML approach, based on a modified version of the well-known dimensionality reduction algorithm SOM \citep{Kohonen}.


\subsection{MLPQNA algorithm}
MLPQNA is a neural network based on the classical Multi-layer Perceptron with two hidden layers \citep{Rosenblatt1963}. Instead of the standard Backpropagation learning rule this algorithm uses a more sophisticated Quasi Newton approximation of the Hessian error matrix \citep{Nocedal2006}. MLPQNA was successfully used for \mbox{photo-z} estimation in a number of papers (such as, \citealt{Biviano2013,Nicastro2018,Brescia2013,Brescia2014,Cavuoti2012,Brescia2012,photoraptor,Cavuoti2016,Cavuoti2017}).\\
The model hyper-parameters were heuristically selected on the basis of our past experience and on an intensive test campaign. To determine the number of neurons in each layer, we follow the rule of thumb described in \citet{Brescia2013}. This rule implies that the optimal number of neurons for the first hidden layer equals $2N+1$ and the optimal number of neurons in the second hidden layer is $N-1$, where $N$ is number of features (i.e., photometric bands). The MLPQNA hyper-parameters are reported in Tab.~\ref{table:MLPQNA_params}.

\begin{table}
\centering
 \begin{tabular}{c c } 
 \hline
 Parameter name & Parameter value \\[0.5ex] 
 \hline\hline
 decay  & 0.001 \\ \hline
 restarts & 80 \\ \hline
 threshold & 0.01 \\ \hline
 epochs & 20000 \\ \hline
 activation function & tanh \\ \hline
 nHiddenLayers & 2 \\ \hline
 number of neurons in the first layer & ($2\cdot$num\_features)+1 \\ \hline
 number of neurons in the second layer & num\_features - 1 \\ \hline
 cost function & Mean squared error \\ \hline
 \end{tabular}
 \caption{MLPQNA model hyper-parameters settings (see \citealt{Brescia2013} for the details).}
 \label{table:MLPQNA_params}
\end{table}


\subsection{Metrics}\label{Metrics}
In order to evaluate the quality of \mbox{photo-z} estimations, we calculate the residuals as: 
\begin{equation}
    \Delta z = \dfrac{z_{\mathrm{spec}}-z_{\mathrm{phot}}}{1+z_{\mathrm{spec}}}
\end{equation}

and apply the following set of statistical metrics:
\begin{enumerate}
    \item Standard deviation $\sigma (\Delta z)$;
    \item Normalized median absolute deviation $\mathrm{NMAD} (\Delta z)=1.48 \times median(|\Delta z|) $, which is less sensitive to catastrophic outliers than~$\sigma (\Delta z)$;
    \item Bias, defined as $mean(\Delta z$);
    \item Percentage of outliers $\eta_{0.15}$, defined as a number of sources with $|\Delta z|\geq0.15$.
\end{enumerate}


\subsection{SOM}\label{SOM}
The Self Organizing Map (SOM) is a neural network algorithm first described in \citet{Kohonen1982}. 
The idea behind SOM is the following: a parameter space of an arbitrary dimensionality is projected on a map of lesser dimensionality, most commonly 2-dimensional, in such a way that neighbour instances in the original parameter space remain neighbours on the resulting map. For this purpose, the SOM algorithm compares the feature vector of every object in the dataset (in our case the magnitude vector of every galaxy) with the weight vector of the same dimensionality associated with each cell on the 2D map. The object is then placed in the cell having the most similar weight vector (in terms of euclidean distance). Such cell is called Best-Matching Unit (BMU). The weights of the BMU and its neighbour cells are updated in such a way that they become more similar to the feature vector of the object. This mechanism ensures that, at the end of training loop, the map learns the representation of the parameter space of the entire training dataset in a self-organized way. That is why SOM is commonly used to perform an unsupervised data exploration. In \mbox{photo-z} estimation applications, several authors demonstrated that SOM can be used for different tasks. For example, \citet{Geach2012} and \citet{Way2012} used it to obtain photo-z, while \citet{Carrasco2014} applied it to estimate \mbox{photo-z} PDFs. Finally, \citet{Masters2015} adapted the SOM to check the coverage of the photometric parameter space by a given spectroscopic sample, thus indirectly creating a suitable method to optimize a spectroscopic follow-up strategy.

In this work we use the SOM for two different purposes: i) to detect potentially unreliable \mbox{spec-z} in the KB (see \mbox{Sect. \ref{methodSOMspectr} and \ref{SOM_spectrZFilter}}); ii)  to ensure that the run dataset occupies the same part of the parameter space as the KB, i.e. to remove objects in the run dataset that are photometrically different from those in the KB (see \mbox{Sect. \ref{methodSOMphotometry} and  \ref{SOM_photoFilter}}). In our experiments we use a modified version of MiniSOM \citep{MiniSOM}\footnote{The original version of MiniSOM can be found in the repository \url{https://github.com/JustGlowing/minisom}. Our modified version is available at \url{https://github.com/ShrRa/minisom}.}.

We use photometric bands as the input features for the SOM training. To analyse the datasets, we colour-label the resulting maps using either \mbox{photo-z}, \mbox{spec-z}, or the number of objects within each cell (also called the cell's occupation). The latter mapping is further referred as occupation maps, and it allows us to check how well the dataset, used for the labelling, covers the parameter space of the SOM training dataset. 
 
 Most of the SOM hyper-parameters (specifically, number of epochs, sigma, learning rate and neighbourhood function) are chosen via grid search. Their final values are listed in Tab.~\ref{table:SOM_params}. The choice of the shape (either square, rectangular or spherical) of the maps is somewhat arbitrary; in a number of previous papers different shapes were tried and proven to be useful (see e.g. \citealt{Buchs2019, Carrasco2014,Masters2015}). We opted for rectangular maps, since it is assumed to give the algorithm a preferred direction to align the data \citet{Masters2015}. It is worth noticing, though, that in the preliminary experiments we did not see any significant effect of the shape of the maps on the analysis, described in the next sections.
 
 The size of the map is chosen in such a way that, on average, each cell contains more than $30$ galaxies from the training set. The choice of this size is based on the best compromise between the two competing goals: the reliability of statistics within each cell, and the need to capture the data topology with the maximum finesse. A SOM with a small number of cells provides us a higher number of galaxies per cell, thus improving the reliability of the statistical indicators. On the other side, a larger SOM shows more details of the data distribution in the parameter space, but the statistics within some cells become unreliable. 
For this reason, we use SOM of different sizes. Specifically, we use small low-resolution maps to determine the anomalous sources within each cell, and large high-resolution maps to investigate the train dataset coverage of the parameter space defined by the run catalogue. 
 
\begin{table}
\centering
 \begin{tabular}{c c } 
 \hline
 Parameter name & Parameter value \\[0.5ex] 
 \hline\hline
 width (low-resolution SOM)  & 25 \\ \hline
 height (low-resolution SOM) & 28 \\ \hline
 width (high-resolution SOM) & 67 \\ \hline
 height (high-resolution SOM) & 64 \\ \hline
 num\_features & 10 \\ \hline
 epochs & 6000 \\ \hline
 sigma & 5 \\ \hline
 learning rate & 0.5 \\ \hline
 neighborhood\_function & bubble \\ \hline
 \end{tabular}
 \caption{SOM settings used for all experiments in this study. For more details about low and high resolution maps, see \S~\ref{SOM}.}
 \label{table:SOM_params}
\end{table}


\subsection{Spec-z cleaning with SOM}\label{methodSOMspectr}

The trained SOM places objects with similar feature vectors in the same or neighbour cells. If some object has a photometry-spec-z relationship that appears to be anomalous for its BMU, such object will be considered as an outlier in terms of \mbox{spec-z} distribution of this BMU. In order to avoid the potential confusion between these in-cell outliers with the traditional outliers of a \mbox{photo-z} distribution, hereafter we will refer the in-cell outliers as \textit{anomalous sources} or \textit{anomalies}. \\
We assume that these objects have a lower \mbox{photo-z} reliability, so we drop them out. In order to exclude such objects, for every galaxy we calculate the coefficient: 
\begin{equation} \label{eq:outlCoeff}
    K_{\mathrm{spec}}=\dfrac{\left<z_{\mathrm{spec}}^{\mathrm{BMU}}\right> - z_{\mathrm{spec}}^{\mathrm{obj}}}{\sigma(z_{\mathrm{spec}}^{\mathrm{BMU}})}
\end{equation}
 where $\left<z_{\mathrm{spec}}^{\mathrm{BMU}}\right>$ is the mean \mbox{spec-z} obtained by averaging over the objects falling in the same BMU, $z_{\mathrm{spec}}^{\mathrm{obj}}$ is the \mbox{spec-z} of the galaxy, and $\sigma(z_{\mathrm{spec}}^{\mathrm{BMU}})$ is the standard deviation of the \mbox{spec-z} distribution within the BMU cell. This coefficient has the same meaning as a standard score, or Z-score, often used in statistics. We choose the prefix $K$ instead of $Z$ to avoid confusion with redshifts.
 Typically, objects are considered to be outliers if $|K_{spec}| >3$. 
 Yet, in our experiments we also tried other criteria to see how it affects the \mbox{photo-z} quality (see \S~\ref{SOM_spectrZFilter}).


\subsection{Parameter space verification with SOM occupation map} \label{methodSOMphotometry}

As we pointed out earlier, neural networks cannot perform extrapolation. It means that, in order to obtain reliable results, the run dataset has to occupy the same area of the parameter space that is well sampled by the KB. \mbox{Fig. \ref{fig:magDistribs}} shows that, in terms of magnitude distribution, our train and run datasets are quite different. 
Therefore, in order to avoid biasing, we need to estimate the statistics for each magnitude bin independently. Furthermore, in order to obtain reliable results, we have to keep the parameter space of the run dataset as similar as possible to the parameter space of the KB. Usually this is done by limiting the magnitudes of the run dataset to the maximum magnitudes of the KB, i.e. by cutting the faint-end "tail" of the magnitude distributions (e.g. \citealt{Cavuoti2015,Wright2019,Eriksen2019}). But this approach poses a major problem: if extreme cuts are adopted, the coverage of the parameter space is ensured but at the risk of loosing many objects with reliable photo-z. With a soft limiting value, on the other hand, the run dataset is affected by too many objects that lay outside of the KB parameter space. 

Using SOM occupation maps, it is possible to implement a more accurate method of photometry filtering, based on the approach first investigated by \citet{Masters2015}. The idea behind the method is simple: we train the SOM on the run dataset and then project the KB on the trained map. Since the entire SOM represents a projection of the parameter space of the run dataset, the galaxies in the KB will be clustered only in a subset of cells, with a fairly large number of cells either poorly occupied or completely empty. For these cells we do not have spectral information and hence \mbox{photo-z} predictions cannot be trusted.
In order to capture in detail the coverage of the run dataset parameter space by the KB, we use a high-resolution SOM map of size $67X64$ (Tab.~\ref{table:SOM_params}).


\section{Experiments} \label{experiments}
The experimental part of this work consists of four stages:
\begin{enumerate}
    \item The first one, described in \S~\ref{expBeforeSOM}, is dedicated to identifying the optimal parameters for MLPQNA. For these experiments, we use our KB with only the standard pre-processing, as outlined in \S~\ref{baseCleanData}.
    \item The second set of experiments is described in \S~\ref{SOM_spectrZFilter}. It aims to analyse how effective is our SOM in removing sources with anomalous \mbox{spec-z}. In this set of experiments we also use only the KB.
    \item At the third stage, discussed in \S~\ref{SOM_photoFilter}, we check whether the SOM can be used to extract from a new dataset (DEIMOS) only those sources that belong to the same part of the parameter space as the KB, i.e. photometrically similar to the sources from the KB. For this set of experiments we use both the KB and the DEIMOS datasets.
    \item Finally, in \S~\ref{Run_Description} we apply the developed methods to estimate the photometric redshifts and their reliability for objects in the run dataset. 
\end{enumerate}


\subsection{Photo-z before SOM cleaning} \label{expBeforeSOM}

Here we calculate the photometric redshifts for the KB after the standard pre-processing described in \S \ref{baseCleanData}. We performed a number of experiments with different combinations of features, namely: all photometric bands, only broad bands, broad bands plus one or more narrow bands, with five "SDSS-like" bands, etc.  The results of the most representative experiments are reported in Tab.~\ref{table:exp_first_run}.

\begin{table*}

\centering
 \begin{tabular}{c c c c c c c c} 
 \hline
 Exp ID & Description & bands & Mean & $\sigma_{\Delta z/(1+z)}$ & NMAD & $\eta_{0.15}$\\ [0.5ex] 
 \hline\hline
 exp001 & SDSS-like, aper2 & \parbox{4.4cm}{u,B,r,ip,zpp}
 & 0.000
 & 0.053
 & 0.025
 & 1.999
 & \\ \hline
 exp002 & SDSS-like, aper3 & \parbox{4.4cm}{u,B,r,ip,zpp}
 & -0.001
 & 0.051
 & 0.027
 & 1.807
 & \\ \hline
 exp003 & SDSS-like, aper\_auto & \parbox{4.4cm}{u,B,r,ip,zpp}
 & -0.003
 & 0.055
 & 0.027
 & 2.364
 & \\ \hline
 exp004 & SDSS-like, aper\_ISO & \parbox{4.4cm}{u,B,r,ip,zpp}
 & -0.004
 & 0.059
 & 0.028
 & 2.346
 & \\ \hline
 exp005 & broad bands, aper2 and aper3 & \parbox{4.4cm}{u,B,r,ip,zpp}
 & -0.001
 & 0.051
 & 0.025
 & 1.756
 & \\ \hline
 exp006 & broad bands, aper3 & \parbox{4.4cm}{B,H,J,Ks,V,Y,ip,r,u,zpp}
 & -0.002
 & 0.048
 & 0.018
 & 2.138
 & \\ \hline
 exp007 & broad bands + one narrow band, aper3 & \parbox{4.4cm}{B,H,J,Ks,V,Y,ip,r,u,zpp,IB574}
 & -0.002
 & 0.048
 & 0.019
 & 1.642
 & \\ \hline
 exp008 & all bands, aper3 & \parbox{4.4cm}{Ks, Y, H, J, B, V, ip, r, u, zp, zpp, IA484, IA527, IA624, IA679, IA738, IA767, IB427, IB464, IB505, IB574, IB709, IB827, NB711, NB816, Hw, Ksw, yH}
 & -0.002
 & 0.049
 & 0.017
 & 1.877
 & \\ \hline
 \end{tabular}
 \caption{Results of the experiments performed with the trained MLPQNA on the blind test set of the KB, after having just applied the standard KB cleaning procedure.}
 \label{table:exp_first_run}
\end{table*}

Keeping in mind that COSMOS2015 has exceptionally rich and well-calibrated photometry, one could expect to obtain \mbox{photo-z} catalogue of high precision. Our expectations were also based on the very precise results obtained with SED fitting in \citet{Laigle2016}, and on the fact that in previous works (e.g. done for SDSS-DR9 \citep{Cavuoti2017} and KiDS \citep{Brescia2014}), MLPQNA generally performed on a comparable or better level than other \mbox{photo-z} algorithms.

Yet, in all the experiments, reported in Tab.~\ref{table:exp_first_run}, we did not reach the accuracy achieved in  previous works. In particular, we have noticeably higher $\sigma (\Delta z_{\mathrm{ML}})$ and percentage of outliers than those reported in the aforementioned papers for KiDS and \mbox{SDSS-DR9}. 

The COSMOS2015 is deeper than the catalogues used in those publications, so the cause of the deterioration of the overall statistics could be in the lower quality of photometry and spectroscopy for the high-z sources. But even for low-z galaxies (i.e. with $z_{\mathrm{spec}}<0.5$), both the percentage of outliers and $\sigma (\Delta z_{\mathrm{ML}})$ turned out to be of lower quality (see Tab.~\ref{table:bin_no_SOM}). In fact, there is an unexpected high percentage of outliers in all redshift bins; they are also uniformly distributed across the whole observed field, meaning that \mbox{photo-z} errors are not caused by the contamination by the light of some nearby star. Also, various feature sets show similar percentage of outliers, so it is unlikely that the source of the problem is the calibration of photometry in some particular band.

\begin{table*}
\centering
  \resizebox{0.8\textwidth}{!}{  

	\begin{tabular}{c r c c r c c c r c}
	\hline
	          & & \multicolumn{4}{c}{ML photo-z} & \multicolumn{4}{c}{\mbox{SED photo-z}}\\
	          \cmidrule(lr){3-6}\cmidrule(lr){7-10}\\
	spec-z bin & Num objects	& $\sigma_{\Delta z/(1+z)}$ & NMAD & Mean~~ & $\eta_{0.15}$ & $\sigma_{\Delta z}$ & NMAD & Mean~~ &  $\eta_{0.15}$ 	\\ [0.5ex] 
	\hline\hline
	[Overall] & 5967 & 0.048 & 0.019 & -0.0023 & 1.64 & 0.094 & 0.011 & -0.0041 & 2.23 \\
	\hline
	[0.0; 0.2] & 422 & 0.100 & 0.027 & -0.0416 & 8.29 & 0.278 & 0.012 & -0.0552 & 7.82 \\
	\hline
    [0.2; 0.4] & 1445 & 0.043 & 0.016 & -0.0081 & 1.31 & 0.056 & 0.008 & -0.0034 & 1.94 \\
    \hline
    [0.4; 0.6] & 1158 & 0.038 & 0.016 & 0.0003 & 1.04 & 0.067 & 0.010 & -0.0002 & 2.07 \\
    \hline
    [0.6; 0.8] & 1489 & 0.036 & 0.018 & 0.0001 & 1.01 & 0.052 & 0.012 & 0.0002 & 1.48 \\
    \hline
    [0.8; 1.0] & 1085 & 0.032 & 0.019 & 0.0041 & 0.55 & 0.044 & 0.014 & 0.0057 & 1.20 \\
    \hline
    [1.0; 1.2] & 368 & 0.051 & 0.028 & 0.0292 & 2.99 & 0.094 & 0.027 & -0.0068 & 3.53 \\
	\hline
	\end{tabular}
    }
\caption{Statistics for the test set calculated in \mbox{spec-z} bins. \mbox{ML photo-z} used here were obtained during the exp007 from Tab.~\ref{table:exp_first_run} using ten broad and one medium bands.}
\label{table:bin_no_SOM}
\end{table*}

Another possible explanation is that the \mbox{photo-z} outliers can be either misinterpreted spectra (in a broad sense of this word, i.e. including blended galaxies, incorrect crossmatch between photometric and spectroscopic catalogues, etc.) or exotic objects. Thanks to the fact that some objects in our main \mbox{spec-z} catalogue contain more than one \mbox{spec-z} measurement, we can estimate how many of the outliers belong to each of these categories. In order to do this, we select objects with more than one measurement of \mbox{spec-z} and calculate maximum difference between these measurements, calling it \mbox{spec-z} scatter. Then we calculate the percentage of outliers separately for objects with a single \mbox{spec-z} measurement, multiple measurements and small $(<0.1)$ scatter, and multiple measurements and large $(\geq 0.1)$ scatter. Tab.~\ref{table:outlAnalysis} reports these calculations for ML and SED fitting photo-z. 

From this table we can deduce several facts: 

\begin{enumerate}
    \item For the objects with multiple \mbox{spec-z} values the percentage of outliers for ML and SED fitting \mbox{photo-z} is essentially the same.
    \item For the objects with small \mbox{spec-z} scatter this percentage is significantly lower ($\eta_{0.15}\sim 0.2\%$) than for the objects with large \mbox{spec-z} scatter ($\eta_{0.15}\sim 11\%$).
    \item For the objects with single \mbox{spec-z} measurement the percentage of outliers for SED fitting is $\sim 50\%$ higher than for ML.
\end{enumerate}

 The most probable explanation here is that for the \mbox{photo-z} outliers with large \mbox{spec-z} scatter, the specific \mbox{spec-z} measurement used to estimate the \mbox{photo-z} residual is incorrect. Consequently, the majority of such bona fide outliers are likely to have correct \mbox{photo-z} predictions. It also implies that a significant percentage of bona fide outliers with single \mbox{spec-z} measurements should be attributed to incorrect \mbox{spec-z} measurements. 
 At the same time, we should not assume that the percentage of incorrect \mbox{spec-z} measurements in this group is the same as for the objects with multiple measurements, since there is no guarantee of similarity of the selection functions for these two categories of objects. 

Instead, we use the SOM cleaning procedure, described in \S~\ref{methodSOMspectr}, to select a set of objects with reliable \mbox{spec-z} even without multiple \mbox{spec-z} measurements.

\begin{table*}
\centering
  \resizebox{0.98\textwidth}{!}{  

    \begin{tabular}{l c c c c c}
	\hline
	          & & \multicolumn{2}{c}{ML photo-z} & \multicolumn{2}{c}{\mbox{SED photo-z}} \\
	          \cmidrule(lr){3-4}\cmidrule(lr){5-6}\\
	Case & Num objects	& Num outliers & \% outliers  & Num outliers & \% outliers \\ [0.5ex] 
	\hline\hline
    Total & 5967 & 98 & 1.64 & 133 & 2.23  \\
    \hline
    Single measurement & 3745 & 63 & 1.68 & 98 & 2.62 \\
    \hline
    \parbox{5cm}{Multiple measurements, \mbox{spec-z} scatter <0.1} & 1945 & 4 & 0.21 & 5 & 0.26\\
    \hline
    \parbox{5cm}{Multiple measurements, \mbox{spec-z} scatter $\geq0.1$} & 277 & 31 & 11.19 & 30 & 10.83\\
    \hline
    \hline
    $|K_\mathrm{spec}| \leq 1$, total & 4311 & 8 & 0.19 & 30 & 0.70\\
    \hline
    $|K_\mathrm{spec}| \leq 1$, single measurement & 2683 & 5 & 0.19 & 24 & 0.89\\
    \hline
    \parbox{6.5cm}{$|K_\mathrm{spec}| \leq 1$, multiple measurements, \mbox{spec-z} scatter <0.1} & 1468 & 0 & 0.00 & 1 & 0.07\\
    \hline
    \parbox{6.5cm}{$|K_\mathrm{spec}| \leq 1$, multiple measurements, \mbox{spec-z} scatter $\geq0.1$} & 160 & 3 & 1.88 & 5 & 3.12\\
        	\hline
	\end{tabular}
    }
\caption{Statistics for ML and SED outliers for the test set of the KB for objects with different number and scatter of \mbox{spec-z} measurements. The upper part of the table reports the results before removing in-cell spec-z anomalies, the lower four rows report the results after $|K_\mathrm{spec}| \leq 1$ filtering.}
\label{table:outlAnalysis}
\end{table*}

\subsection{Photo-z statistics after removing in-cell anomalous spec-z}\label{SOM_spectrZFilter}

\begin{figure*}
    \includegraphics[width=0.98\textwidth]{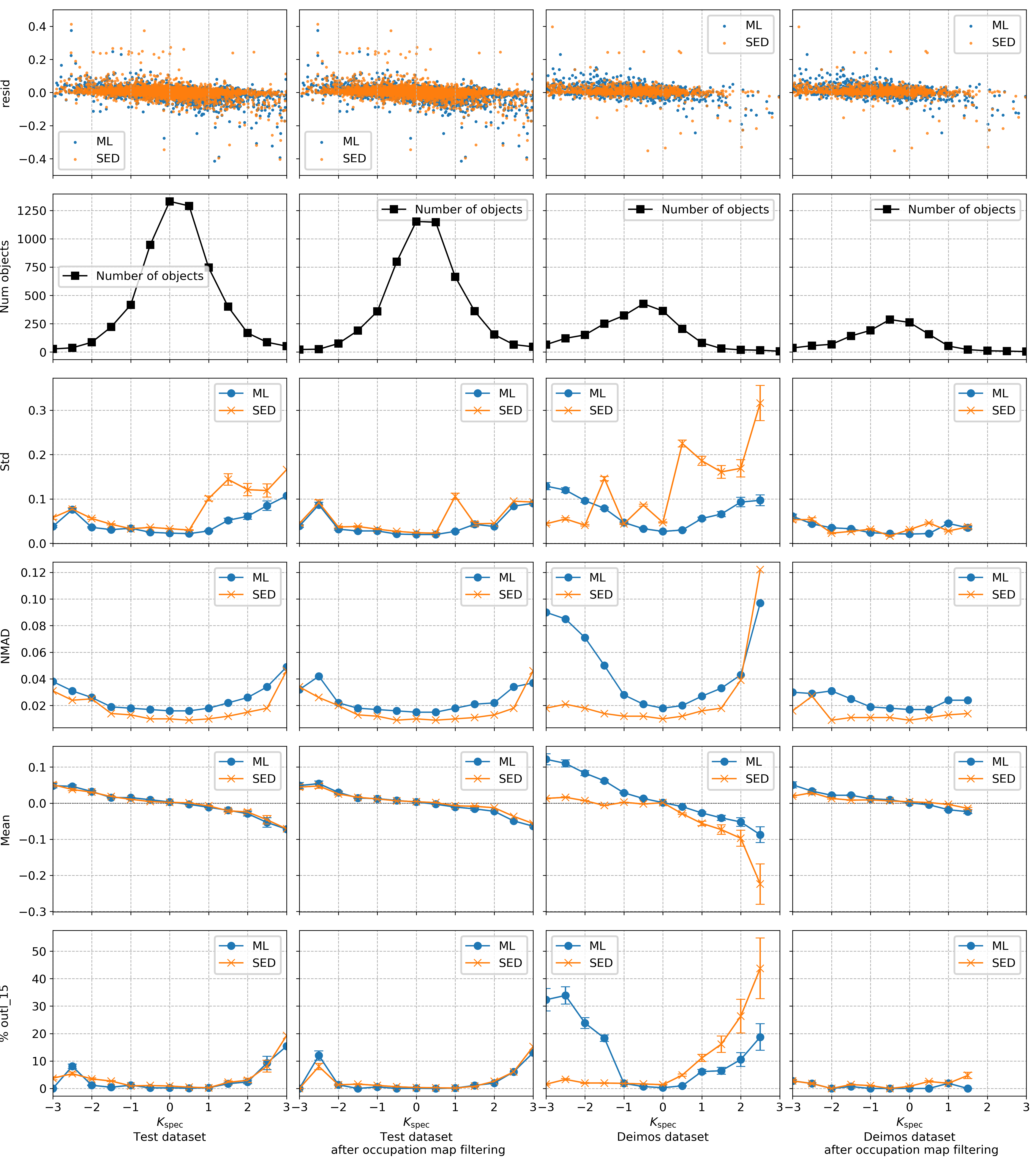}
     \caption{Statistics for the test and DEIMOS datasets in $K_\mathrm{spec}$ bins before and after occupation map filtering. Bins with number of objects < 15 are considered to be unreliable and excluded from these plots.}
     \label{fig:statBinsTestDeimosSpecZOutl}
\end{figure*}

\begin{figure*}
    \includegraphics[height=0.9\textheight]{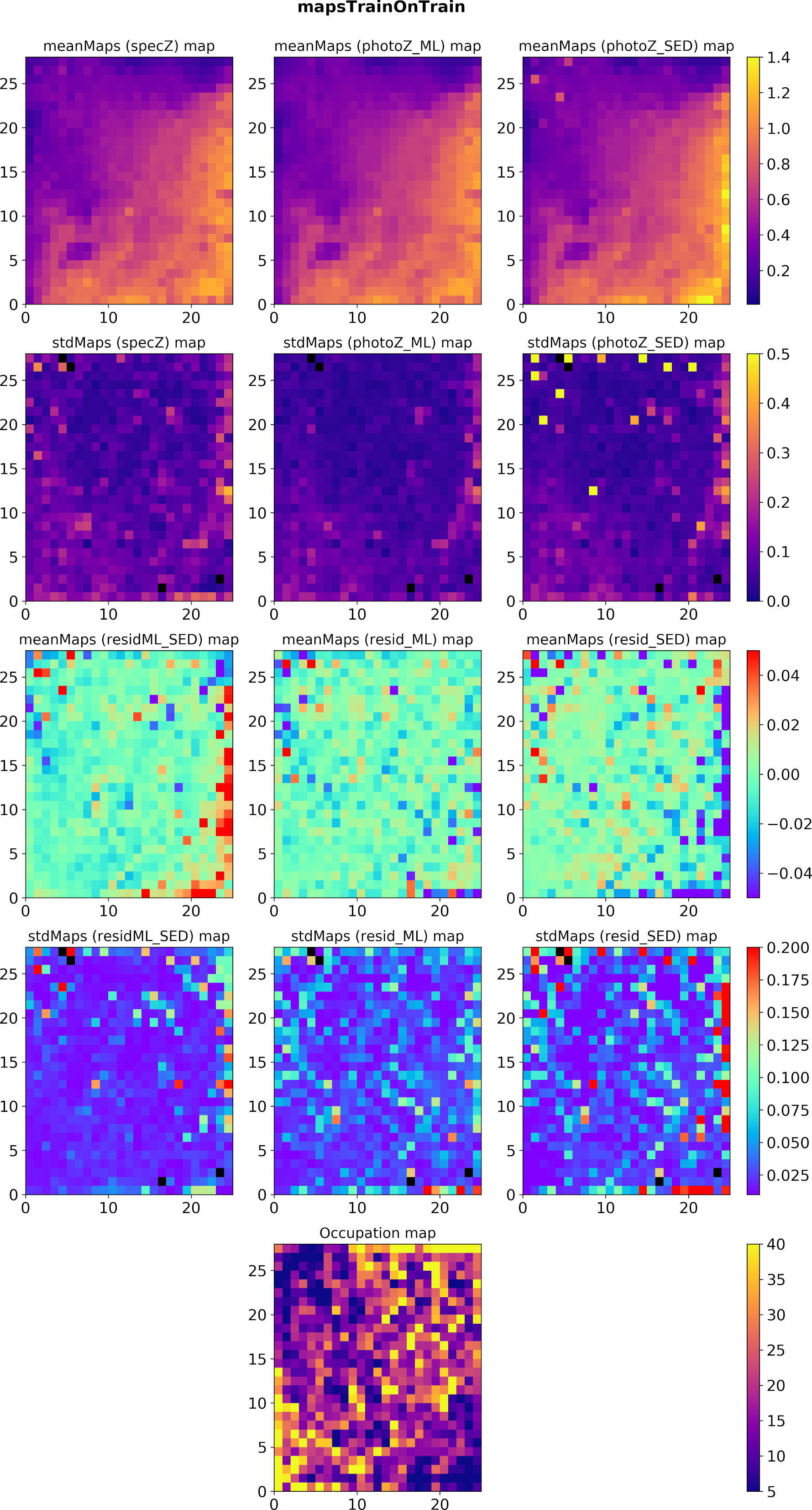}
    \caption{SOMs built and labelled with KB. The first two rows from the top illustrate mean and standard deviation of the redshifts (spectroscopic on the left, ML in the center and SED fitting on the right) within each cell. The third and fourth rows illustrate mean and standard deviation of residuals: on the left are the maps calculated for the residuals between ML and SED fitting photo-z, in the center and on the right are the residuals for ML and SED fitting photo-z. The last plot reflects how many objects are within each cell. Black cells on the occupation and mean maps imply that these cells are empty. On the maps of standard deviation black cells mean that occupation of the cell equals 1 and standard deviation cannot be calculated. } 
    \label{fig:trainOnTrainMaps}
\end{figure*}

We train the SOM, using the KB with the set of broad bands that gave us the best results for the MLPQNA experiments (\texttt{u,B,V,r,ip,zpp,Y,J,H,Ks}, see Tab.~\ref{table:exp_first_run}). This feature set can be viewed as an approximation of the feature set that will exist for the combined LSST and Euclid data, which will also range from optical to NIR wavelengths \citep{Rhodes2017}. \mbox{Fig. \ref{fig:trainOnTrainMaps}} shows the resulting SOM map, color-labelled with, respectively, the mean and standard deviation of \mbox{spec-z} and ML and SED fitting photo-z. In order to discard objects with anomalous \mbox{spec-z} values, we calculate the in-cell \mbox{spec-z} outlier coefficients $K_\mathrm{spec}$ for each source, as defined in  Eq.~(\ref{eq:outlCoeff}); then we bin the whole KB according to the value of this coefficient and calculate the statistics of \mbox{photo-z} residuals for each bin. This allows us to check how the quality of \mbox{photo-z} correlates with similarity between the \mbox{spec-z} of a given source and the mean \mbox{spec-z} of its BMU.\\ 
\mbox{Fig. \ref{fig:statBinsTestDeimosSpecZOutl}} shows that the majority of the objects have relatively small $K_\mathrm{spec}$ (second row of the figure), and the statistics are much better for them than for the objects with larger absolute values of $K_\mathrm{spec}$.\\
In the majority of the bins \mbox{ML photo-z}'s have lower standard deviations (third row from the top) and lower percentage of outliers (last row from the top) than \mbox{SED photo-z}'s, but higher NMAD (fourth row). Predictably, mean residuals (second row from the bottom) have inverse correlation with $K_\mathrm{spec}$, implying that \mbox{photo-z} predictions, for  objects with \mbox{spec-z} lower than the median \mbox{spec-z} of their BMU, are biased towards higher values, and vice versa.\\ 
By limiting our dataset to galaxies with absolute value of $K_\mathrm{spec}$ smaller than $1$, we reduce the percentage of outliers from $1.64$ to $0.19$ for \mbox{ML photo-z} and from $2.23$ to $0.7$ for SED fitting. The standard deviation of residuals also is reduced by a factor $\sim 2$ (see Tab.~\ref{table:testAfterCleanings}).\\ Remarkably, after this cleaning the statistics are effectively the same as for the objects with multiple \mbox{spec-z} measurements and low \mbox{spec-z} scatter. In other words, removing in-cell anomalous \mbox{spec-z} leaves us with a reliable set of \mbox{spec-z} with no need to use repeated \mbox{spec-z} measurements, and this set is twice as large as the one with multiple measurements.

The lower part of Tab.~\ref{table:outlAnalysis} shows that while the percentage of outliers for the \mbox{ML photo-z} drops not only for the objects with large \mbox{spec-z} scatter, but for other categories as well, for the SED fitting the improvement for the objects with single \mbox{spec-z} measurement is weaker.\\

Taking into account that the run dataset does not contain spec-z, it appeared useful to check whether  the quality of the dataset could be improved by using \mbox{photo-z} in-cell anomaly detection instead of \mbox{spec-z}. To do so, we calculated outlier coefficients for SED and \mbox{ML photo-z} ($K_\mathrm{SED}$ and $K_\mathrm{ML}$ correspondingly), and applied the same filtering using these coefficients instead of $K_\mathrm{spec}$. It turns out that such filtering improves the overall statistics, even though less than the $K_\mathrm{spec}$ cleaning (upper half of Tab.~\ref{table:testAfterCleanings}). As expected, the relative improvements are stronger for the SED fitting \mbox{photo-z} than for the ML photo-z; in \S~\ref{discuss_PhotoZfilter} we will discuss this aspect in better detail.

\begin{table*}
\centering
  \resizebox{\textwidth}{!}{  

	\begin{tabular}{l c c c c c c c c c}
	\hline
	          & & \multicolumn{4}{c}{ML photo-z} & \multicolumn{4}{c}{\mbox{SED photo-z}}\\
	          \cmidrule(lr){3-6}\cmidrule(lr){7-10}\\
	Filtering & Num objects	& $\sigma_{\Delta z}$ & NMAD & Mean &  $\eta_{0.15}$  & $\sigma_{\Delta z}$ & NMAD & Mean & $\eta_{0.15}$ 	\\ [0.5ex] 
	\hline\hline
    No filtering & 5967 & 0.048 & 0.019 & -0.0023 & 1.64 & 0.094 & 0.011 & -0.0041 &   2.23  \\
    	\hline
    $|K_\mathrm{spec}| \leq 1$ & 4311 & 0.025 & 0.017 &  ~0.0002 & 0.19 & 0.052 & 0.010 &  ~0.0006 &   0.70  \\
    	\hline
    $|K_\mathrm{ML}| \leq1$ & 4071 & 0.045 & 0.018 & -0.0020 & 1.28 & 0.077 & 0.010 & -0.0026 &   1.65  \\
    	\hline
    $|K_\mathrm{SED}| \leq1$ & 4133 & 0.043 & 0.018 & -0.0020 & 1.06 & 0.061 & 0.010 & -0.0021 &   1.43  \\
    	\hline\hline	
	trainMapOccupation $>5$ & 5167 & 0.041 & 0.018 & -0.0021 & 1.18 & 0.058 & 0.010 &  ~0.0001 &   1.49  \\
    	\hline
	$K_\mathrm{spec}$ + trainMapOccupation & 3761 & 0.022 & 0.016 & -0.0002 & 0.05 & 0.05 & 0.009 &  ~0.0018 &   0.35  \\
    	\hline
	$K_\mathrm{ML}$ + trainMapOccupation & 3587 & 0.039 & 0.017 & -0.0015 & 1.06 & 0.062 & 0.010 & -0.0003 &   1.17  \\
    	\hline
	$K_\mathrm{SED}$ + trainMapOccupation & 3624 & 0.038 & 0.017 & -0.0016 & 0.94 & 0.038 & 0.010 &  ~0.0002 &   0.99  \\
    	\hline
	\end{tabular}
    }
\caption{Statistics for ML and SED fitting \mbox{photo-z} calculated for the test dataset after different types of filtering. The upper part of the table presents the statistics calculated for the dataset without \mbox{spec-z} and \mbox{photo-z} outliers. The lower part reports the effects of photometric filtering with occupation map.}
\label{table:testAfterCleanings}
\end{table*}

\subsection{Photo-z after parameter space verification with occupation map}\label{SOM_photoFilter}

\begin{figure*}
    \includegraphics[width=0.7\textwidth]{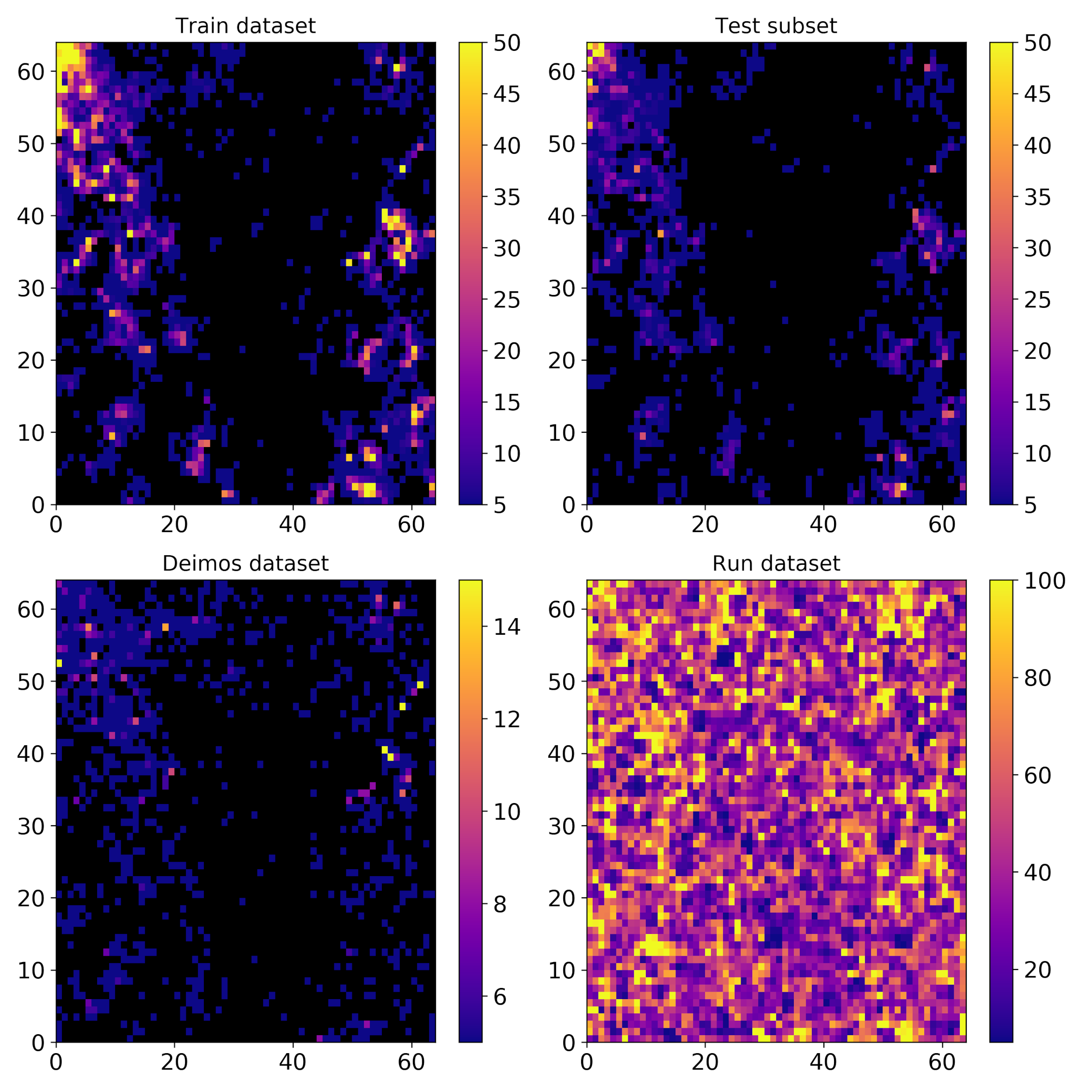}
    \caption{Occupation maps for all datasets projected on SOM trained with run dataset.}
    \label{fig:activMaps}
\end{figure*}

\begin{figure*}
    \includegraphics[width=0.98\textwidth]{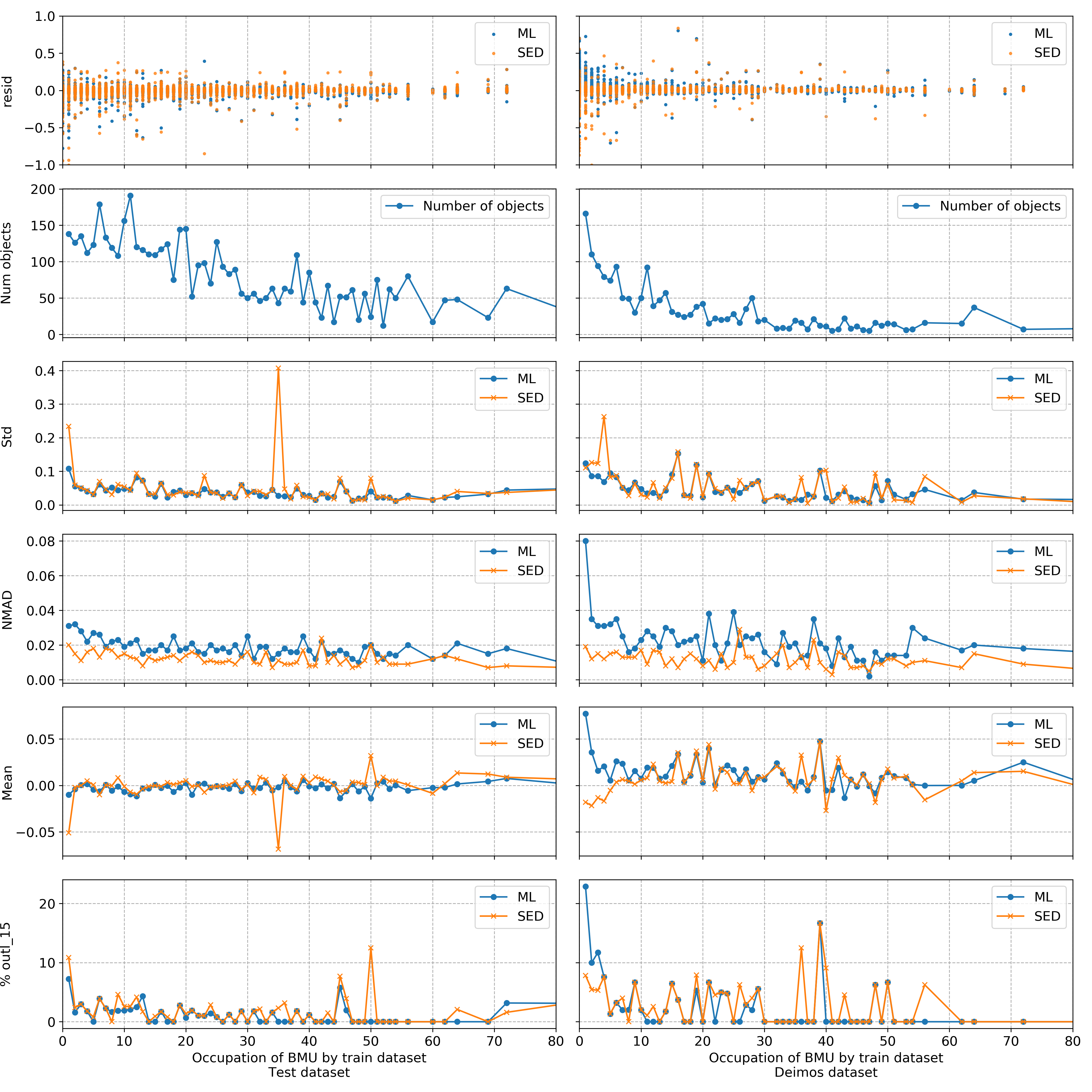}
     \caption{Dependency of the statistics for the test and DEIMOS datasets from the occupation of their BMU by objects from train dataset. The x-axis is limited to occupation $\leq80$ since there are not enough objects in the cells with bigger occupations to calculate reliable statistics.}
     \label{fig:statOccupationTestDeimos}
\end{figure*}

In order to be able to discard the part of the run dataset for which MLPQNA predictions are not likely to be reliable, we have to use occupation maps to identify and flag galaxies that are not photometrically similar to the galaxies of the train dataset. To estimate the performance of such photometry cleaning, we made use of the DEIMOS dataset. Considering that the DEIMOS dataset is slightly deeper than the train dataset, we expect that the cleaning process will mostly remove faint objects. 
\mbox{Fig. \ref{fig:activMaps}} shows occupation maps for the train, test, DEIMOS and run datasets, respectively. For each object of the test, DEIMOS and run datasets we derive the occupation of its BMU by the galaxies of the train dataset. As expected, \mbox{Fig. \ref{fig:statOccupationTestDeimos}} demonstrates that the statistics tend to be better for the cells with higher occupation by the train dataset. Based on the statistics for the DEIMOS dataset (right panel of \mbox{Fig. \ref{fig:statOccupationTestDeimos}}), we chose to leave only the objects that belong to the cells with occupation $>5$, since more strict criteria do not bring further improvements.

\begin{figure*}
    \includegraphics[height=0.85\textheight]{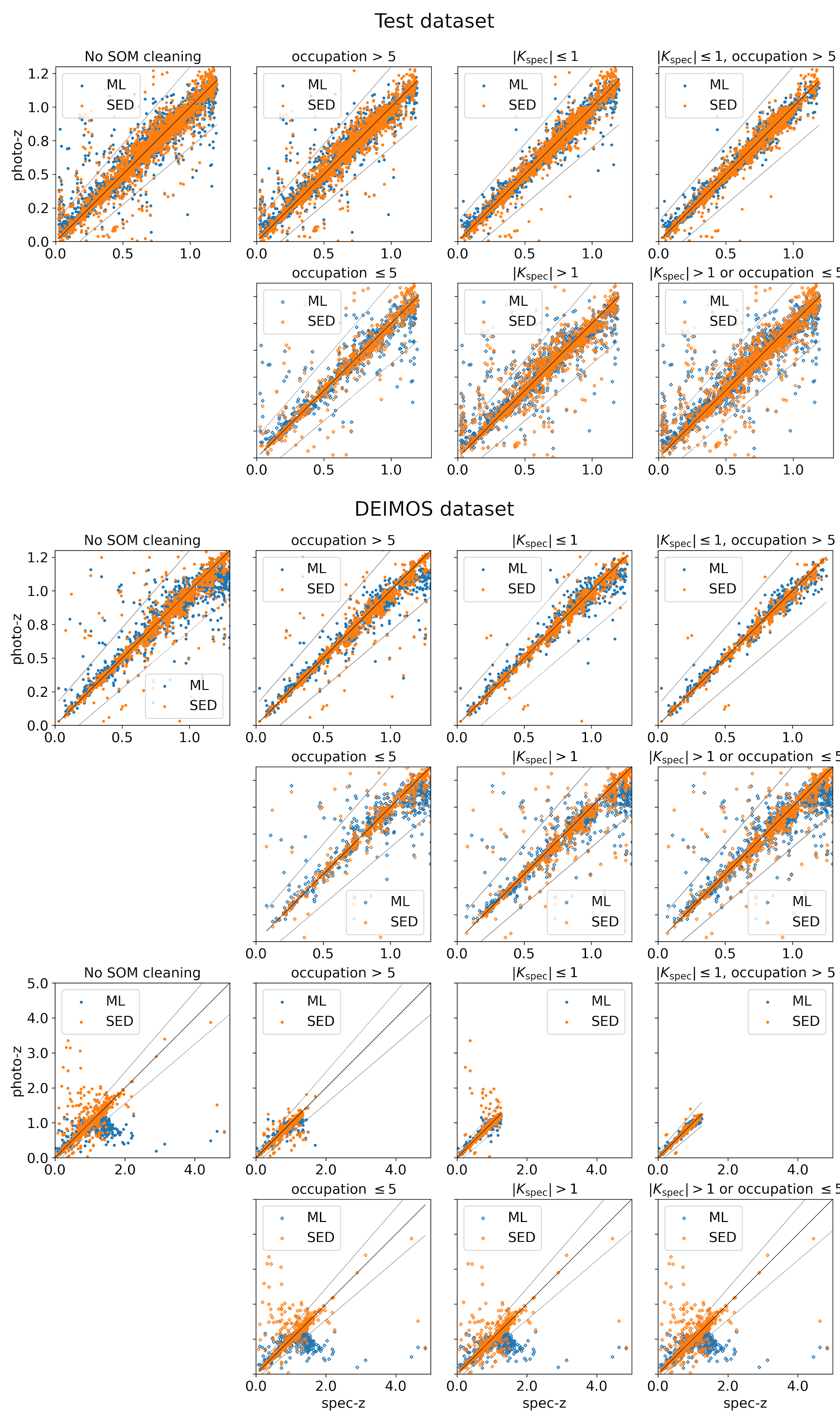}
     \caption{Scatter plots of ML and SED fitting $z_{\mathrm{phot}}$ against $z_{\mathrm{spec}}$. In the first column on the left are the datasets before SOM filtering procedures, in the second column are the results after only $z_{\mathrm{spec}}$ outlier filtering, in the third are the plots after only occupation map filtering, and in the last column are the datasets after the two cleaning together. The dotted lines show outlier boundaries defined as $z_{\mathrm{photo}}=z_{\mathrm{spec}} \pm 0.15\cdot(1+z_{\mathrm{spec}})$.
     First two rows from the top correspond to the test dataset, the next two rows report the results for the DEIMOS dataset, limited by $z_{spec}<1.2$ and $z_{phot}<1.2$, the last two rows demonstrate the full DEIMOS dataset ($z_{spec<5}$, $z_{phot<5}$). In every pair of rows the first row shows the galaxies that remain in the dataset after the filtering, and the second row shows the galaxies that have been removed.} 
     \label{fig:TestScatter}
\end{figure*}

Tables \ref{table:testAfterCleanings} and \ref{table:DeimosAfterCleanings} report the overall statistics for the test and DEIMOS datasets, respectively. The statistics are given for the different stages of cleaning, including the cases of combination of cleaning with occupation map and removal of the objects with anomalous spectroscopic or photometric redshifts. From these tables we see that the effect of occupation filtering is much stronger for the DEIMOS than for the test dataset (since the test dataset by construction covers the same regions of the parameter space as the train dataset). Still, even on the test dataset the effect of the filtering with occupation map is comparable or better than the effect of the filtering of the \mbox{photo-z} anomalous sources. Also, occupation map filtering is more cost-effective in a sense that it discards much less objects than \mbox{spec-z} or \mbox{photo-z} anomalous source filtering.\\
From Tab.~\ref{table:DeimosAfterCleanings} we see that without any filtering applied to the DEIMOS  dataset, our \mbox{\mbox{ML photo-z}} has worse indicators than \mbox{\mbox{SED photo-z}}. This can be easily understood by remembering that part of the DEIMOS dataset lays outside of the parameter space of the train dataset. But for objects that belong to the cells with good occupation, the percentage of outliers and standard deviations for both ML and SED fitting \mbox{photo-z} are very close.\\
After additional \mbox{spec-z} anomalous source filtering, the statistics improves even more and reaches approximately the same values as for the KB; the scatter plots in the bottom part of Fig.~\ref{fig:TestScatter} clearly illustrate the improvement. From there, it can be seen that the two cleaning procedures remove different outliers. The plots for the DEIMOS dataset show that both ML and SED fitting photo-z require occupation map filtering to select the objects with good predictions. Without it, SED fitting produces a lot of catastrophic outliers in the whole range of $z_{\mathrm{spec}}$, and ML systematically fails for the objects with $z_{\mathrm{spec}}>1.2$. $K_{spec}$ filtering eliminates most of the remaining outliers, distributed randomly across the $z_{spec}$ range.

\begin{table*}
\centering
  \resizebox{\textwidth}{!}{  

	\begin{tabular}{l c c c c c c c c c}
	\hline
	          & & \multicolumn{4}{c}{ML photo-z} & \multicolumn{4}{c}{\mbox{SED photo-z}}\\
	          \cmidrule(lr){3-6}\cmidrule(lr){7-10}\\
	Filtering & Num objects	& $\sigma_{\Delta z}$ & NMAD & Mean & $\eta_{0.15}$  & $\sigma_{\Delta z}$ & NMAD & Mean &  $\eta_{0.15}$ 	\\ [0.5ex] 
	\hline\hline
    No filtering & 2255 & 0.099 & 0.032 & 0.0347 & 10.86 & 0.142 & 0.014 & -0.0082 & 5.06  \\
        	\hline
    $|K_\mathrm{spec}| \leq 1$ & 1075 & 0.035 & 0.020 & 0.0018 & ~1.02 & 0.127 & 0.011 & -0.0103 & 2.88  \\
        	\hline
    $|K_\mathrm{ML}| \leq 1$ & 1209 & 0.095 & 0.033 & 0.0392 & 12.57 & 0.090 & 0.013 & -0.0031 & 4.55  \\
        	\hline
    $|K_\mathrm{SED}| \leq 1$ & 1183 & 0.085 & 0.029 & 0.0245 & ~8.96 & 0.078 & 0.013 & ~0.0005 & 3.30  \\
        	\hline\hline
    trainMapOccupation>5 & 1382 & 0.058 & 0.023 & 0.0127 & ~2.10 & 0.059 & 0.012 & ~0.0085 & 2.68  \\
        	\hline
    $K_\mathrm{spec}$ + trainMapOccupation & ~758 & 0.025 & 0.018 & 0.0017 & ~0.13 & ~0.031 & 0.010 & 0.0038 & 0.92  \\
        	\hline
    $K_\mathrm{ML}$ + trainMapOccupation & ~724 & 0.064 & 0.022 & 0.0136 & ~1.93 & 0.060 & 0.011 & 0.0080 & 2.62  \\
        	\hline
    $K_\mathrm{SED}$ + trainMapOccupation & ~741 & 0.046 & 0.020 & 0.0063 & ~1.48 & 0.044 & 0.011 & ~0.0075 & 1.89  \\
        	\hline
	\end{tabular}
    }
\caption{Statistics for ML and SED fitting \mbox{photo-z} for the DEIMOS dataset after different types of filterings. Upper part of the table describes statistics after \mbox{spec-z} and \mbox{photo-z} outlier removal, while the lower part reports the effects of photometry filtering with occupation map.}
\label{table:DeimosAfterCleanings}
\end{table*}

\subsection{SOM cleaning of the run dataset}\label{Run_Description}

The COSMOS2015 contains $\sim500\,000$ galaxies. After the standard pre-processing described in \S~\ref{baseCleanData} and excluding objects that belong to the KB, the run catalogue consists of $\sim190\,000$ galaxies. For all these galaxies we calculate \mbox{ML photo-z}, and in order to determine the reliability of these redshifts, we calculate $K_{\mathrm{SED}}$, $K_{\mathrm{ML}}$ and the occupation of their BMU by the objects in the train dataset, as described in the previous subsections. 

Since we do not have \mbox{spec-z} for the galaxies in the run  dataset, we can only perform an accuracy test comparing SED and \mbox{ML photo-z}. To do so, we construct ML/SED residuals $\Delta z_{\mathrm{ML/SED}} = (z_{\mathrm{SED}}-z_{\mathrm{ML}})/(1+z_{\mathrm{SED}})$.  Obviously, such a test suffers from biases introduced by both \mbox{photo-z} methods, so it can be used only for qualitative estimation of the \mbox{photo-z} robustness.

Tab.~\ref{table:RunAfterCleanings} compares the statistics for \mbox{ML/SED photo-z} residuals for the DEIMOS and run catalogues after applying different filters. The cleaning with occupation map appears to be the most important step, since it allows to reduce the percentage of outliers by almost an order of magnitude. Removal of SED fitting \mbox{photo-z} anomalous sources also improves the statistics. Maximum improvement for both datasets is achieved by a combination of occupation map filtering and removal of SED fitting \mbox{photo-z} anomalous sources. It reduces the percentage of outliers from~11\% to 0.27\% for the DEIMOS dataset, and from~40\% to ~2\% for the run dataset. $\sigma(\Delta z)$ for the DEIMOS drops from $0.102$ to $0.03$, and from $0.238$ to $0.056$ for the run dataset. NMAD changes from $0.033$ to $0.022$ for the DEIMOS dataset and from $0.129$ to $0.026$ for the run dataset.

\begin{table*}
\centering
  \resizebox{\textwidth}{!}{  

	\begin{tabular}{l c c c c c c c c c c}
	\hline
	          & \multicolumn{5}{c}{DEIMOS dataset} & \multicolumn{5}{c}{Run dataset}\\
	          \cmidrule(lr){2-6}\cmidrule(lr){7-11}\\
	Filtering & Num objects	 & $\sigma_{\Delta z}$ & NMAD & Mean & $\eta_{0.15}$  & Num objects & $\sigma_{\Delta z}$ & NMAD & Mean &  $\eta_{0.15}$ 	\\ [0.5ex] 
	\hline\hline
    No filtering & 2255 & 0.102 & 0.033 & ~0.0355 & 11.35 & 194509 & 0.238 & 0.129 & ~0.1373 & 40.23  \\
    			\hline
    $|K_\mathrm{spec}| \leq 1$ & 1075 & 0.070 & 0.023 & ~0.0049 & ~2.88 & NA & NA & NA & NA & NA   \\
    			\hline
    $|K_\mathrm{ML}| \leq 1$ & 1209 & 0.088 & 0.034 & ~0.0392 & 11.75 & 137152 & 0.225 & 0.110 & ~0.1243 & 37.76  \\
    			\hline
    $|K_\mathrm{SED}| \leq 1$ & 1183 & 0.072 & 0.030 & ~0.0224 & ~7.44 & 146773 & 0.206 & 0.101 & ~0.1217 & 34.93  \\
    			\hline\hline	
    trainMapOccupation>5 & 1382 & 0.047 & 0.023 & ~0.0031 & ~1.52 & ~43279 & 0.083 & 0.028 & -0.0019 & ~3.59  \\
    			\hline
    $K_\mathrm{spec} $ + trainMapOccupation & ~758 & 0.036 & 0.020 & -0.0028 & ~0.92 & NA & NA & NA & NA & NA   \\
    			\hline
    $K_\mathrm{ML} $ + trainMapOccupation & ~724 & 0.043 & 0.022 & ~0.0048 & ~0.97 & ~33296 & 0.080 & 0.028 & -0.0029 & ~3.48  \\
    			\hline
    $K_\mathrm{SED}$ + trainMapOccupation & ~741 & 0.030 & 0.022 & -0.0015 & ~0.27 & ~35189 & 0.056 & 0.026 & -0.0031 & ~2.17  \\
            	\hline
	\end{tabular}
    }
\caption{Statistics for ML/SED residuals for the DEIMOS and run datasets after different types of filtering. Upper part of the table describes statistics after \mbox{spec-z} and \mbox{photo-z} outlier removal, lower part reports the effects of photometry filtering with occupation map. Taking into account that for the run dataset spectral information is absent, the rows corresponding to \mbox{spec-z} cleanings for run dataset are empty.}
\label{table:RunAfterCleanings}
\end{table*}

\subsection{Purely data-driven spec-z sample}\label{DataDrivenOnly}

Apart from selecting reliable photo-z predictions from the final catalogue, SOM filtering can be used to create a larger and potentially more representative KB. It can be done by including in the KB not only galaxies with 'good' spectral quality flags (see \S~\ref{spec-z_prep}), but also those galaxies that have 'non-robust' quality flags, but good SOM quality indicator ($|K_{spec}|\leq1$), i.e. those that appear to have typical $z_{spec}$ for their color cell. In practice, it means that we omit the \texttt{Q\_f} cleaning and use only $|K_{spec}|\leq1$ filtering. The KB constructed in this exclusively data-driven way consists of $24\,058$ objects instead of $19\,893$.

For this sample, we calculated photo-z using the same MLPQNA setup as in the previous sections. For additional comparison, we calculated photo-z for a random galaxy sample of the same size, taken from the non-cleaned spec-z dataset. Tab.~\ref{table:dirty_specZ} reports the results of these experiments.

From this table it can be seen that by using only SOM in-cell anomaly filtering, we are unable to obtain the same quality of photo-z as if we make preliminary selection using spectral quality flags. However, for the ML photo-z the difference is rather small; with the combined standard and SOM cleaning, we have $\mathrm{NMAD} = 0.017$ and $\eta_{0.15} = 0.19$, while with SOM-only cleaning $\mathrm{NMAD} = 0.018$ and $\eta_{0.15} = 0.56$; the mean residual is essentially the same for both cases. This deterioration of statistics comes with a benefit of increasing the size of the final spec-z dataset by $\sim 67\%$. SED fitting photo-z show slightly stronger degradation: while NMAD remains of the same order, the mean bias is worsened by two orders, from $0.0006$ to $-0.0134$, and the percentage of outliers increases from 0.7 to 1.93. The experiment with a random sample taken from the uncleaned spec-z catalogue demonstrates that without any form of data cleaning, photo-z quality is significantly worse.

This purely data-driven selection of reliable spec-z can be suitable not only for photo-z prediction, where larger KB means better representation of the run dataset, but also for a general verification and validation of the quality of spec-z catalogues, useful in other contexts.

\begin{table*}
\centering
  \resizebox{\textwidth}{!}{  

	\begin{tabular}{l l l l l r l l l r l}
	\hline
	          & & & \multicolumn{4}{c}{ML photo-z} & \multicolumn{4}{c}{\mbox{SED photo-z}}\\
	          \cmidrule(lr){4-7}\cmidrule(lr){8-11}\\
	Exp ID & Description & Num obj	& $\sigma_{\Delta z}$ & NMAD & Mean & $\eta_{0.15}$  & $\sigma_{\Delta z}$ & NMAD & Mean &  $\eta_{0.15}$ 	\\ [0.5ex] 
	\hline\hline
 exp007 & Baseline & 5967 & 0.048 & 0.019 & -0.0023 & 1.64 & 0.094 & 0.011 & -0.0041 &   2.23 \\\hline
 
 exp007-clean & Baseline, $K_\mathrm{spec}\leq 1$ & 4311 & 0.025 & 0.017 &  0.0002 & 0.19 & 0.052 & 0.010 &  0.0006 &   0.7 \\\hline
 
 d\_exp027 & 'Dirty' spec-z, $|K_\mathrm{spec}|\leq 1$ &  7218 &   0.033 &    0.018 &  -0.0009 &          0.57 &    0.135 &     0.012 &   -0.0134 &           1.93 \\\hline 
 d\_exp028 & 'Dirty' spec-z, random sample & 7218 &   0.072 &    0.025 &  -0.0055 &          4.32 &    0.178 &     0.015 &   -0.0238 &           5.78 \\\hline
 \end{tabular}
 }
 \caption[Photo-z performance for the 'dirty' spec-z sample]{Photo-z performance for the test sample with and without standard \texttt{Q\_f} cleaning. The first two rows describe baseline \texttt{exp007} experiment, which was done on the spec-z sample cleaned with standard \texttt{Q\_f} criteria (\S~\ref{spec-z_prep}), before and after SOM in-cell spec-z anomaly filtering. The third row describes an experiment done on the spec-z sample that was not cleaned with \texttt{Q\_f}, but only with SOM in-cell filtering. The last row reports an experiment done on a random sample from the spec-z catalogue which was not cleaned neither with \texttt{Q\_f} nor SOM in-cell filtering.}
 \label{table:dirty_specZ}
\end{table*}

\section{Discussion} \label{discussion}

As we have shown in the previous section, there are several ways in which we can apply SOM to improve the quality of the \mbox{photo-z} catalogues. In some cases these cleaning procedures improve both the standard deviation of residuals and the percentage of outliers (e.g.~Tab.~\ref{table:DeimosAfterCleanings}). 
In this section we compare the effects of the SOM cleaning on ML and SED fitting \mbox{photo-z} (\S~\ref{discuss_SEDvsML}), consider the nature of \mbox{spec-z} and \mbox{photo-z} in-cell anomalies (\S~\ref{discuss_SpecZfilter} and \S~\ref{discuss_PhotoZfilter}), and define general strategies for using the SOM cleaning methodology on other datasets (\S~\ref{discuss_strategy}).

\subsection{SED fitting vs. ML} \label{discuss_SEDvsML}

In all the experiments performed on the KB, the \mbox{ML photo-z} distribution has a lower percentages of outliers and lower standard deviations than the SED fitting \mbox{photo-z} distribution, but higher NMAD (see Tab.~\ref{table:testAfterCleanings}). For the DEIMOS dataset, before SOM filtering, the situation is different: \mbox{ML photo-z} have a significantly higher percentage of outliers ($\sim 11\%$  against 5\% for SED fitting), due to the fact that  DEIMOS contains many objects laying outside of the boundaries of the parameter space sampled by the KB. Occupation map filtering discards the majority of these outliers and the statistics of both ML and SED fitting \mbox{photo-z} become similar to those for the test dataset (see Tab.~\ref{table:testAfterCleanings} and \ref{table:DeimosAfterCleanings}). In particular, for ML $\sigma_{\Delta z}$ decreases from $0.099$ to $0.058$ (for the KB it equals $0.041$), and for SED fitting $\sigma_{\Delta z}$  from $0.142$ to $0.059$ ($0.058$ for the KB). The percentage of outliers also drops from $10.86\%$ to $2.1\%$ for \mbox{ML photo-z} ($1.18\%$ for the KB) and from $5.06\%$ to $2.68\%$ for SED fitting ($1.49\%$ for the KB).

The difference between the two methods in terms of NMAD can be explained by the fact that SED fitting methods benefit from the inclusion of the narrow-band photometry. It allows the SED fitting methods to detect emission lines passing through a certain wavelength range \citep{Ilbert2009}. On the contrary, in the case of MLPQNA, the same inclusion of the additional narrow bands does not lead to significant improvements (Tab.~\ref{table:exp_first_run}). Earlier work by several authors (see e.g. \citealt{Heinis2016} and \citealt{PAUS}) demonstrated that inclusion of the narrow-band photometry generally improves \mbox{ML photo-z} only after additional preparations, such as transfer learning or a preliminary extensive phase of feature selection. A possible explanation to this is that in comparison to the broad bands, narrow bands carry redshift signal for a smaller fraction of galaxies, namely, only for those galaxies whose spectrum has emission or absorption lines passing through a particular narrow band. In order to learn to extract this signal, a ML model might need a much larger training set than when only broad bands are used.

Even after occupation map filtering and removing in-cell anomalous spec-z, SED fitting produces a higher number of catastrophic outliers than ML. Also, the residuals for these outliers are larger than for ML outliers (see Fig.~\ref{fig:TestScatter}; for the test dataset the mean absolute value of residuals for SED fitting catastrophic outliers is 0.35, while for ML outliers it's 0.26; for the DEIMOS the mean absolute residual for SED fitting is 0.29, and for ML it's 0.22). As can be seen from the second row of Fig.~\ref{fig:trainOnTrainMaps}, the SOM map of the standard deviations of SED fitting \mbox{photo-z} has a number of cells with significantly higher values than appear on the $\sigma(z_\mathrm{spec})$ and $\sigma(z_\mathrm{ML})$ maps. These are the cells where the majority of the catastrophic outliers are located, and it seems likely that these outliers appear due to the lack of a suitable SED template or SED fitting failure.

An unexpected result is that SED fitting seriously benefits from occupation map filtering. For the test dataset, the percentage of outliers drops from 2.23\% to 1.49\%, while for the DEIMOS it goes from 5.06\% to 2.68\%, which is even better than what is obtained after \mbox{spec-z} filtering. One possible explanation to this is that the objects in the areas of parameter space that are poorly covered by the \mbox{spec-z} catalogue are likely to be faint and at a high spec-z, thus having a less reliable \mbox{photo-z} due either to the lack of suitable SED templates or to the photometry issues. 

The aforementioned results for \mbox{ML photo-z} are obtained using only 10 broad and 1 medium band. In the case of SED fitting \mbox{photo-z} more than thirty (broad, medium and narrow) bands were used; still, the statistics for SED fitting and ML are quite close. It implies some consequences for the observational strategies of the future surveys. Depending on the design of a survey, it can be more beneficial to invest resources into obtaining either a larger spectroscopic KB with \mbox{ML photo-z} algorithms in mind, or additional medium/narrow band photometry for SED fitting. Both decisions lead to similar quality of the \mbox{photo-z} catalogue with some differences in the NMAD and in the percentage of outliers. Obviously, obtaining both an extensive spectroscopic catalogue and medium/narrow band photometry will allow to use both techniques and to increase the reliability of the \mbox{photo-z} predictions (e.g.~\citealt{Cavuoti2017}, also see \S~\ref{discuss_strategy}).

\subsection{Spec-z anomalous sources filtering} \label{discuss_SpecZfilter}

Removal of the \mbox{spec-z} anomalous sources drastically reduces the percentage of outliers for both \mbox{photo-z} methods. As we showed in \S~\ref{expBeforeSOM}, the \mbox{photo-z} outliers mostly appear to be misinterpreted spectra, and removing them from the analysis makes the performance estimation more correct. Yet, there are much more objects with high values of $K_{spec}$ than ML or SED fitting outliers. A question arises, what are all these objects.

Looking at Fig.~\ref{fig:statBinsTestDeimosSpecZOutl}, we can see that in general the statistics (including NMAD, which is less sensitive to hard outliers) deteriorate smoothly with the increase of the absolute value of $K_\mathrm{spec}$. This smooth dependency is present not only for \mbox{ML photo-z}, but for SED fitting \mbox{photo-z} as well. It means that within each cell of SOM (which represents a small hyper-volume in the photometric parameter space), both ML and SED fitting perform better for "typical" \mbox{spec-z} than for atypical ones. This is true even for those regions of parameter space that are well-covered by the \mbox{spec-z} catalogue. In other words, $K_{spec}$ works as an additional, finer indicator of whether a given galaxy is well-represented in the spectroscopic KB, although it does not offer any insight on the reasons of why this galaxy is not well-represented - whether it is due to the physical rarity of its type, to the catalogues' selection function or other reasons. 

In general, removing objects with high $K_\mathrm{spec}$ allows us to create a reliable \mbox{spec-z} sample, useful for comparing the performance of different \mbox{photo-z} algorithms. In our case, for the test dataset this sample is more than two times larger than the set of the objects with multiple \mbox{spec-z} measurements that are in good agreement with each other. A purely data-driven approach to the selection of reliable spec-z, described in \S~\ref{DataDrivenOnly}, allows us to enlarge this sample even more, with a comparatively small deterioration of the quality of the data. In every case, the SOM cleaning mostly does not change the overall shape of the redshift distribution of the samples, with an exception of slightly stronger filtering in the lowest and highest redshift bins, where the percentage of outliers is the highest (see Fig.~\ref{fig:specZDistribs} and Fig.~\ref{fig:specZDistribDirty} in Appendix~\ref{AppendA}). Consequently, the usage of this method seems to be highly beneficial for the preparation of any future compound \mbox{spec-z} catalogues, and possibly for the verification of the new \mbox{spec-z} surveys against the old ones. 

\subsection{Photo-z anomalous sources and occupation map filtering} \label{discuss_PhotoZfilter}

To clean  \mbox{photo-z} values for the run dataset, the only possibility is to use $K_{\mathrm{ML}}$ or $K_{\mathrm{SED}}$ to discard either ML or SED fitting \mbox{photo-z} anomalies. However, the performance is different in the two cases. As it can be seen from Tab.~\ref{table:testAfterCleanings}
and \ref{table:DeimosAfterCleanings}, discarding \mbox{SED fitting} anomalies improves the statistics slightly more than removing ML anomalous sources. Obviously, this is related to the large percentage of outliers obtained in  SED fitting \mbox{photo-z} and to their large residuals. 

On the DEIMOS dataset, \mbox{photo-z} anomalous sources filtering, together with occupation map cleaning allows to reach $\sigma_{\mathrm{\Delta z}} \approx 0.046$ and percentage of outliers $\approx 1.48 \%$ for \mbox{ML photo-z}, and $\sigma_{\mathrm{\Delta z}} \approx 0.044$ and percentage of outliers $\approx 1.89 \%$ for SED fitting.  At the same time, for ML/SED residuals on the DEIMOS the same cleaning procedure brings $\sigma_{\mathrm{\Delta z}} \approx 0.03$ and percentage of outliers $\approx 0.27\%$, which is quite close to the statistics for ML and SED photo-z after occupation map filtering together with removing spec-z anomalous sources. 

\begin{figure*}
    \includegraphics[width=0.98\textwidth]{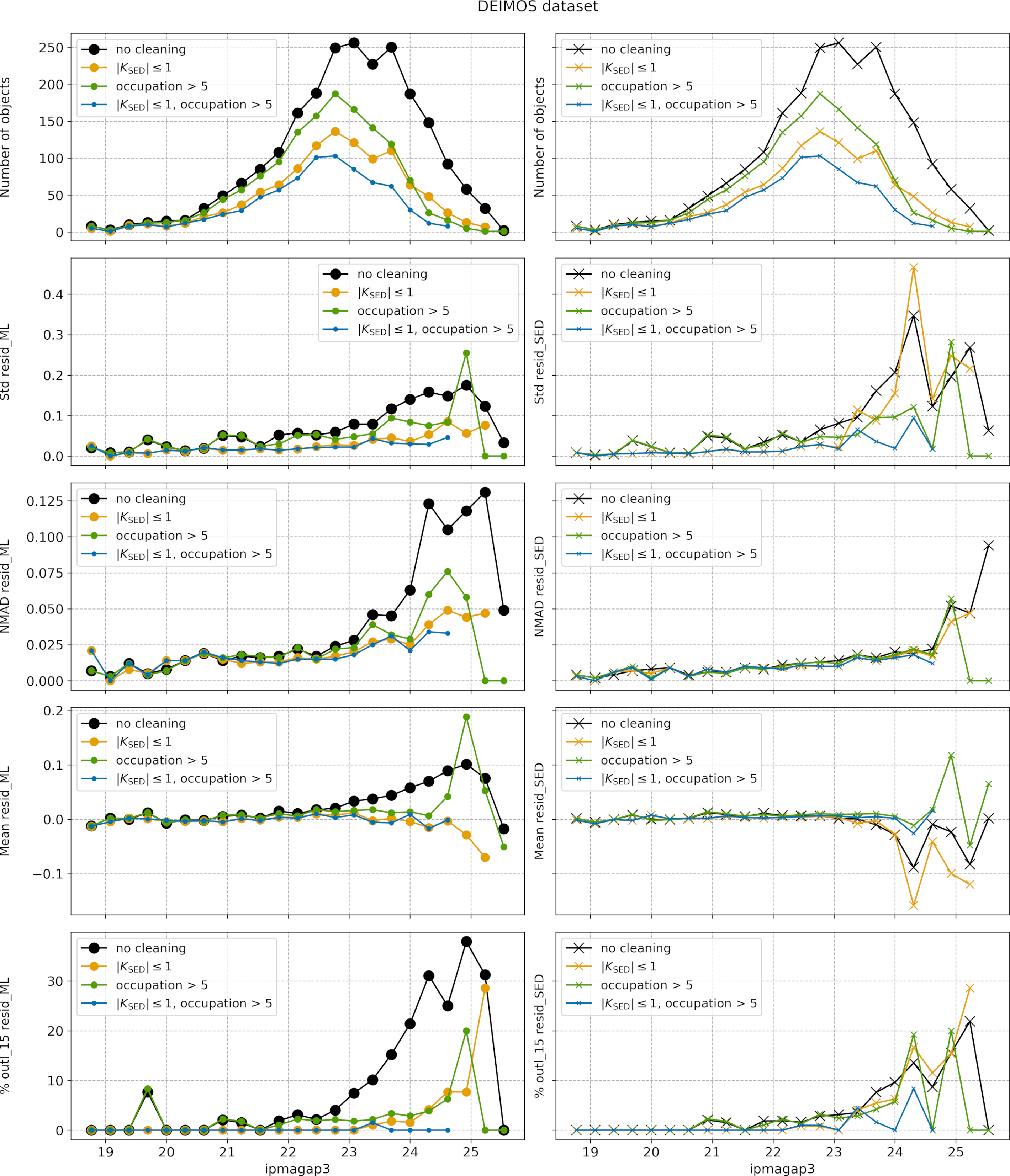}
     \caption{Statistics for the DEIMOS datasets in \texttt{ipmagap3} bins after applying different filters. Left panel: \mbox{ML photo-z} residuals, right panel: SED fitting \mbox{photo-z} residuals. Bins with number of objects < 15 are considered to be unreliable and excluded from these plots.}
     \label{fig:statMagBinsDeimosAfterCleaning}
\end{figure*}

\begin{figure*}
    \includegraphics[height=0.9\textheight]{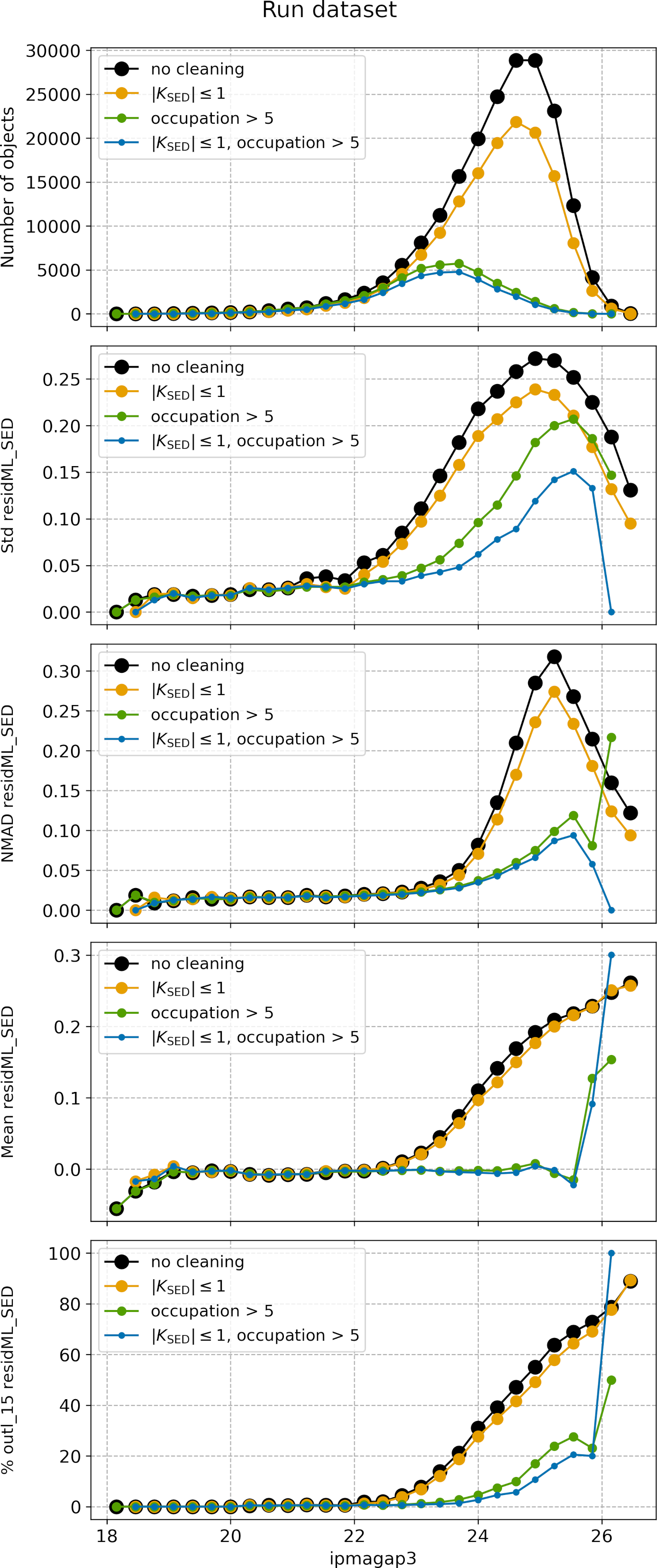}
     \caption{Statistics for ML/SED residuals for run dataset in \texttt{ipmagap3} bins after applying different filters. Bins with number of objects < 15 are considered to be unreliable and excluded from these plots.}
     \label{fig:statMagBinsRunAfterCleaning}
\end{figure*}

In \mbox{Fig. \ref{fig:statMagBinsDeimosAfterCleaning}} we plot the values of different statistics in \texttt{ipmagap3} magnitude bins at different stages of cleaning of the DEIMOS dataset. As expected, with occupation map cleaning the improvements are mostly achieved due to the filtering in the fainter part of magnitude distribution. Instead of completely loosing the faint objects, as it would have happened if we used the traditional cut-off procedure, we are preserving those which have fairly good quality of \mbox{photo-z} predictions. \mbox{Fig. \ref{fig:statMagBinsRunAfterCleaning}} shows the same picture for the run dataset. For what the standard deviation is concerned, the difference appears for objects fainter than \mbox{\texttt{ipmagap3}~$\sim 21$}, but the most significant effect is observed for objects with \mbox{\texttt{ipmagap3}~$> 23.$}

\subsection{Filtering strategies} \label{discuss_strategy}

The optimal strategy of the data cleaning with SOM depends on the nature of a dataset and on the task at hand:
\begin{itemize}
    \item Identification of \mbox{spec-z} anomalous sources is useful for finding unreliable \mbox{spec-z} and under-represented objects. 
    \item Identification of \mbox{photo-z} anomalies allows us to improve the quality of a \mbox{photo-z} catalogue, especially for SED fitting \mbox{photo-z}. It can be applied to the run catalogue (i.e., for objects without spectral information), but this procedure is effective only for datasets that are well sampled by the KB.
    \item For the run datasets that are not well sampled by the KB, SOM filtering with an occupation map is the most important step. It leads to significant improvements not only in the case of ML based methods but also in the case of SED fitting.
\end{itemize}

In this work we use all three types of cleaning with the following thresholds:
\begin{enumerate}
    \item \mbox{spec-z} anomalies filtering: $K_\mathrm{spec} \leq 1$.
    \item Occupation map filtering: $\texttt{trainMapOccupation} > 5$.
    \item \mbox{photo-z} anomalies filtering: $K_\mathrm{SED} \leq 1$.
    \item In cases when better NMAD is needed and percentage of outliers is less critical, SED fitting \mbox{photo-z} are preferable. For the tasks that demand the lowest percentage of outliers, \mbox{ML photo-z} show better results. Finally, it is possible to choose between the two values of photo-z: when \mbox{photo-z} predictions are similar (based on the grid search on the test and DEIMOS datasets we recommend to use the criteria of \mbox{ML/SED residual $< 0.5$}), it is better to select SED fitting value, and when the residual is $> 0.5$, \mbox{ML photo-z} are more likely to be correct. Tab.~\ref{table:selectPhotoZ} reports the statistics for \mbox{photo-z} selected in this way for the test and DEIMOS datasets. With such selection we obtain both good NMAD and low percentage of outliers. 
\end{enumerate}

\begin{table*}
\centering
  \resizebox{0.9\textwidth}{!}{  

	\begin{tabular}{l c c c c c c c c c c}
	\hline
	          & \multicolumn{5}{c}{Test dataset} & \multicolumn{5}{c}{DEIMOS dataset}\\
	          \cmidrule(lr){2-6}\cmidrule(lr){7-11}\\
	Photo-z & Num objects	 & $\sigma_{\Delta z}$ & NMAD & Mean &  $\eta_{0.15}$  & Num objects & $\sigma_{\Delta z}$ & NMAD & Mean &  $\eta_{0.15}$ 	\\ [0.5ex] 
	\hline\hline
    ML & 3508 & 0.021 & 0.016 & -0.0001	& 0.03 & 725 & 0.024 & 0.018 & 0.0020	& 0  \\
    			\hline
    SED & 3508 & 0.049 & 0.009 & -0.0001 & 0.06 & 725 & 0.027  & 0.009 & 0.0022 & 0.55  \\
    			\hline
    Mixed & 3508 & 0.018 & 0.009 & ~0.0012 & 0.03 & 725 & 0.017 & 0.010 & 0.0051 & 0\\
    			\hline			
	\end{tabular}
    }
\caption{Statistics for ML, SED fitting and mixed (selected based on the ML/SED residual value) residuals for the test and DEIMOS datasets. Both datasets were cleaned with the following conditions: $\mathrm{trainMapOccupation}>5$, $|K_\mathrm{spec}| \leq 1$, $|K_\mathrm{SED}| \leq 1$.}
\label{table:selectPhotoZ}
\end{table*}

\section{Conclusions} \label{conclusions}

In this work, we calculated \mbox{ML photo-z} for the COSMOS2015 catalogue using MLPQNA algorithm. For the training and testing we used multi-instrument spectroscopic KB with $z_{\mathrm{spec}} \leq 1.2$ and various sets of photometric bands, obtaining the best results with a feature set composed by 10 broad and 1 narrow bands. The comparison of the statistics for \mbox{ML photo-z} and SED fitting photo-z, calculated by \citealt{Laigle2016} using the whole set of COSMOS2015 bands, showed that \mbox{ML photo-z} has lower percentage of outliers and $\sigma_{\Delta z}$, but higher NMAD. Particularly, for the test dataset without additional SOM cleaning MLPQNA produces \mbox{photo-z} with $\sigma_{\Delta z}=0.048$, $\mathrm{NMAD}=0.019$ and $ \eta_{0.15} = 1.64\%$, while SED fitting \mbox{photo-z} has $\sigma_{\Delta z}=0.094$, $\mathrm{NMAD}=0.011$ and $\eta_{0.15} = 2.23\%$ (Tab.~\ref{table:testAfterCleanings}).

We found that the \mbox{ML photo-z} algorithm does not benefit from the inclusion of the most medium and narrow photometric bands (Tab.~\ref{table:exp_first_run}). Finding a way to exploit the information contained in those bands should be a subject of further work.

Our experiments demonstrated that a significant percentage of outliers have similar values of ML and SED fitting photo-z. By analysing the objects with multiple \mbox{spec-z} measurements we discovered that the majority of such outliers have unreliable \mbox{spec-z} values, which makes us to believe that the \mbox{photo-z} prediction for these objects are correct and the actual percentage of outliers for both \mbox{photo-z} methods is significantly lower. On the subset of galaxies with multiple similar \mbox{spec-z} measurements, $\eta_{0.15} \sim 0.2\%$ for both \mbox{photo-z} methods.\\

We tested the possibility of using Self-Organizing Maps (SOM) for removing unreliable spec-z and creating a high-quality \mbox{spec-z} sample. To do this, we calculated a coefficient $K_\mathrm{spec}$ that quantifies how much a spectrum of a given galaxy differs from the mean spectra of all the galaxies belonging to the same SOM cell, e.g. of the galaxies that are most photometrically similar. We found that $K_\mathrm{spec}$ allows us to remove objects with incorrect spec-z: the percentage of outliers for the test set of the KB drops from $1.64$ to $0.19$ for \mbox{ML photo-z} and from $2.23$ to $0.7$ for SED fitting, and $\sigma_{\Delta z}$ for both ML and SED fitting \mbox{photo-z} improves almost by a factor of $\sim 2$ (see Tab.~\ref{table:testAfterCleanings}). At the same time, $K_\mathrm{spec}$ is sensitive to the intrinsic in-homogeneity of the galaxy population, caused by physical reasons. In this way $K_\mathrm{spec}$ serves as a fine indicator of whether a given object is well-represented within the KB.

To ensure that our run dataset occupies the same area of the parameter space as the KB, we also used SOM following the methodology that we call \textit{occupation map} cleaning. On the control DEIMOS \mbox{spec-z} dataset, that is slightly deeper than our KB \mbox{spec-z} catalogue,  we found that by using this cleaning, we reduce the percentage of outliers from ~11\% to ~2\% for \mbox{ML photo-z} and from ~5\% to ~3\% for SED fitting photo-z, with consequent improvements of other metrics. The details are reported in Tab.~\ref{table:DeimosAfterCleanings}.

We find that we can also improve the statistics by excluding the objects with SED fitting \mbox{photo-z} values that are anomalous for their SOM cells. Removing objects with anomalous \mbox{ML photo-z} does not improve the results significantly. Using both occupation map and SED fitting \mbox{photo-z} in-cell anomalies filtering, we are able to bring the statistics for the DEIMOS dataset to the order of those for the KB. To be more precise, the standard deviation drops from $\sigma_{\Delta z}=0.099$ to $\sigma_{\Delta z}=0.046$ for ML and from $\sigma_{\Delta z}=0.142$ to $\sigma_{\Delta z}=0.044$ for SED fitting, and the percentage of outliers lessens from $\eta_{0.15}=10.86$ to $\eta_{0.15}=1.48$ for ML and from $\eta_{0.15}=5.06$ to $\eta_{0.15}=1.89$ for SED fitting. This result allows us to select the parts of photometric catalogues for which our \mbox{photo-z} predictions, obtained with any algorithm, can be trusted.

All in all, the SOM in-cell anomaly detection, presented in this work, proved to be a viable method for selecting reliable \mbox{spec-z} samples from a contaminated catalogue and a good tool for identifying SED fitting \mbox{photo-z} outliers. The SOM occupation map filtering also seems to be recommendable for ensuring the reliability of the future \mbox{photo-z} catalogues. We plan to investigate the potential of these methods in application to the other kinds of the astronomical datasets in the future works.


\section{Data availability}
The source catalogues for this work can be obtained as described in Sect.~\ref{baseCleanData}. The final catalogue, containing MLPQNA \mbox{photo-z} and SOM-produced parameters that can be used for selecting objects with high-confidence predictions will be published via CDS Vizier facility. The code for reproducing this work is available in the GitHub repository \url{https://github.com/ShrRa/COSMOS_SOM}. The MLPQNA software is available within the PhotoRApToR\footnote{\url{http://dame.oacn.inaf.it/dame_photoz.html\#photoraptor}}  (PHOTOmetric Research APplication To Redshifts \citealt{photoraptor}) package.

\section*{Acknowledgements}
We are grateful to the anonymous referee for the suggestions and comments that helped to improve this paper. 

This work has received financial support from the European Union’s Horizon 2020 research and innovation program under the Marie Skłodowska-Curie grant agreement No. 721463 to the SUNDIAL ITN network. Stefano Cavuoti acknowledges financial support from FFABR 2017 (Fondo di Finanziamento per le Attività Base di Ricerca). OR thanks Valeria Amaro, Civita Vellucci, Maurizio D'Addona and Kseniia Sysoliatina for useful discussions and technical help at different stages of this research. MB acknowledges financial contributions from the agreement \textit{ASI/INAF 2018-23-HH.0, Euclid ESA mission - Phase D} and the \textit{INAF PRIN-SKA 2017 program 1.05.01.88.04}.



\bibliographystyle{mnras}
\bibliography{references} 



\onecolumn
\clearpage
\appendix
\renewcommand{\thefigure}{A. \arabic{figure}}
\setcounter{figure}{0}

    \section{Additional information on data processing and its effects on the redshift distributions}\label{AppendA}
    \clearpage
    \vspace*{0.1cm}
    \begin{figure*}
     \centering
     \includegraphics[width=0.9\textwidth]{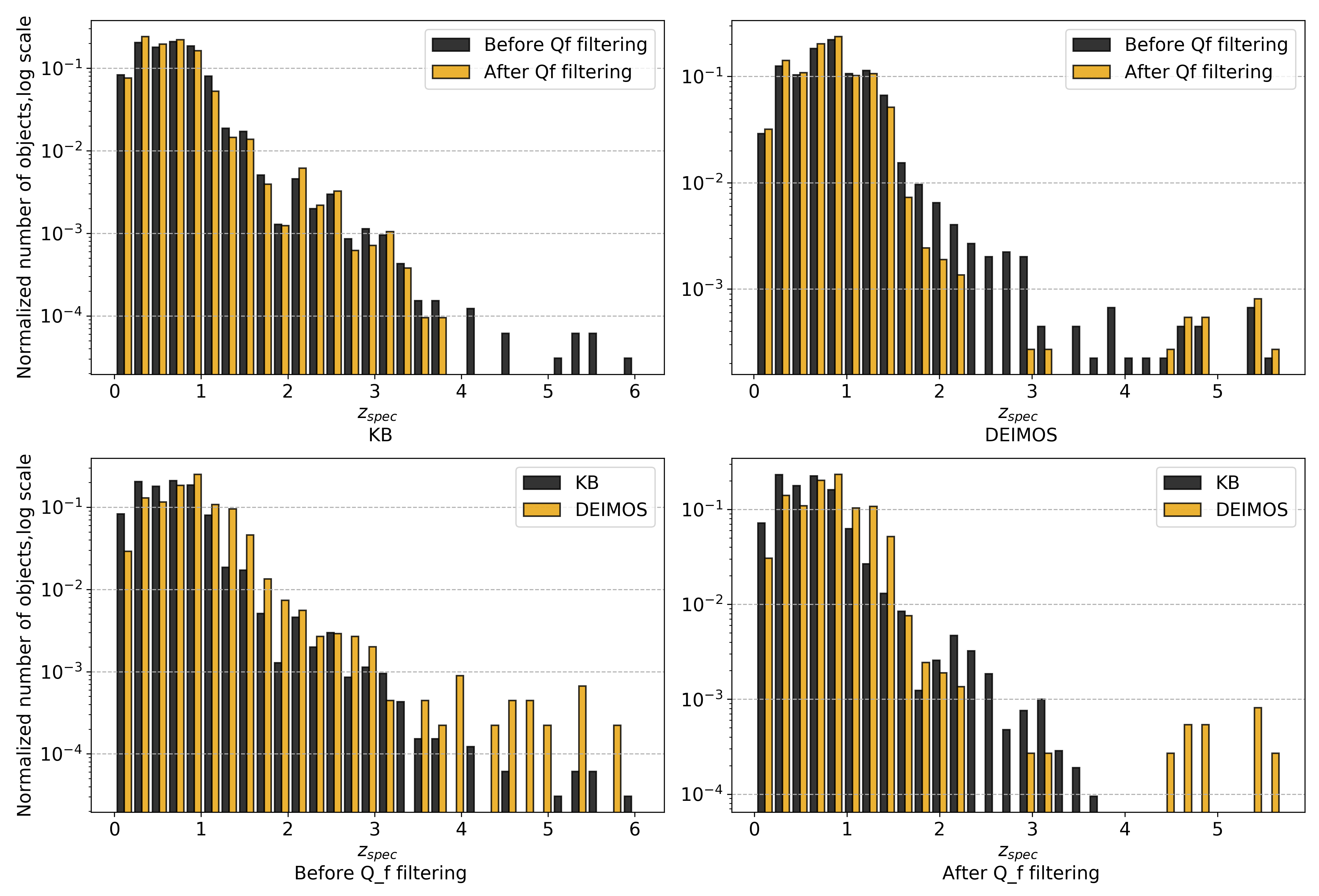}
     \caption{Normalized spec-z distribution before and after quality flag (\texttt{Q\_f}) cleaning of the KB and DEIMOS datasets. The plots in the top row compare distributions dataset-wise, while the plots in the bottom row compare the KB and DEIMOS on the same cleaning stages.}
     \label{fig:spectrZ-Qf}
    \end{figure*}
    
    \begin{figure*}
    \centering
     \includegraphics[width=0.98\textwidth]{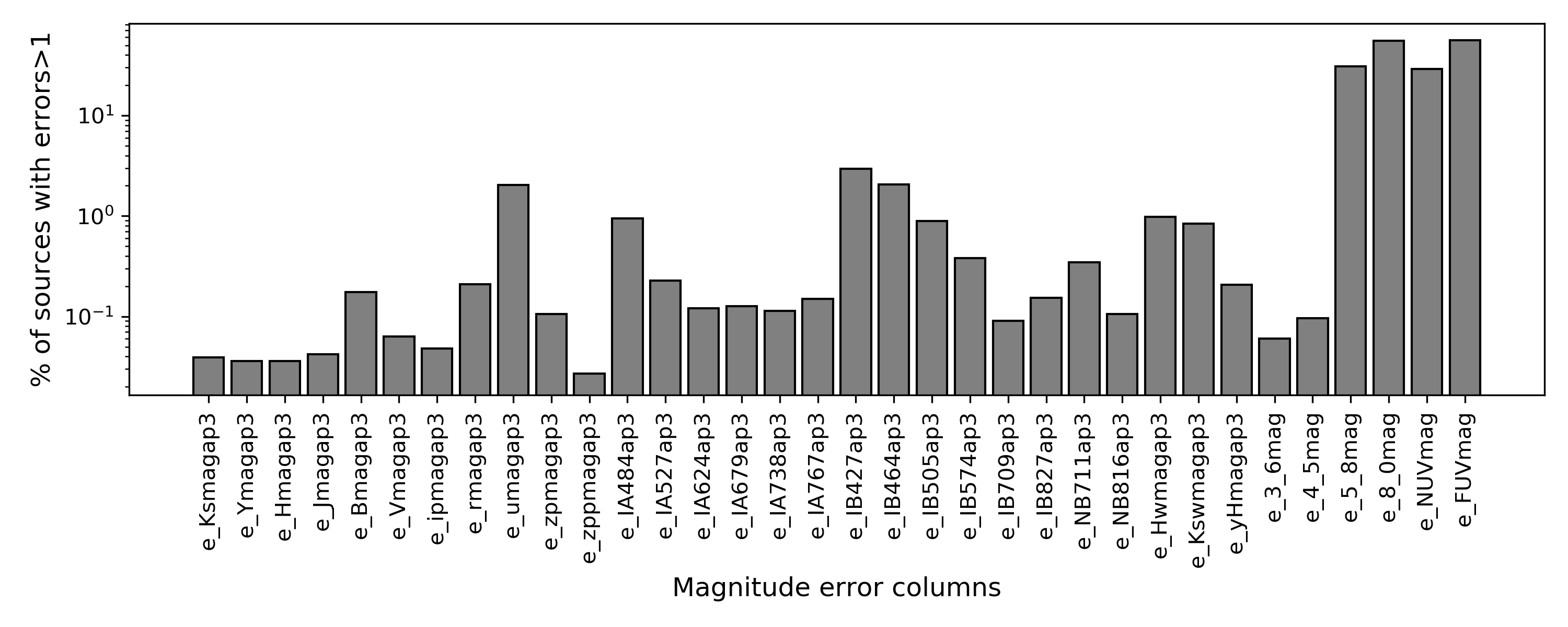}
     \caption{Number of objects in the KB with magnitude errors>1 by bands.}
     \label{fig:magErr}
    \end{figure*}
    
    \begin{figure*}
    \centering
     \includegraphics[width=0.9\textwidth]{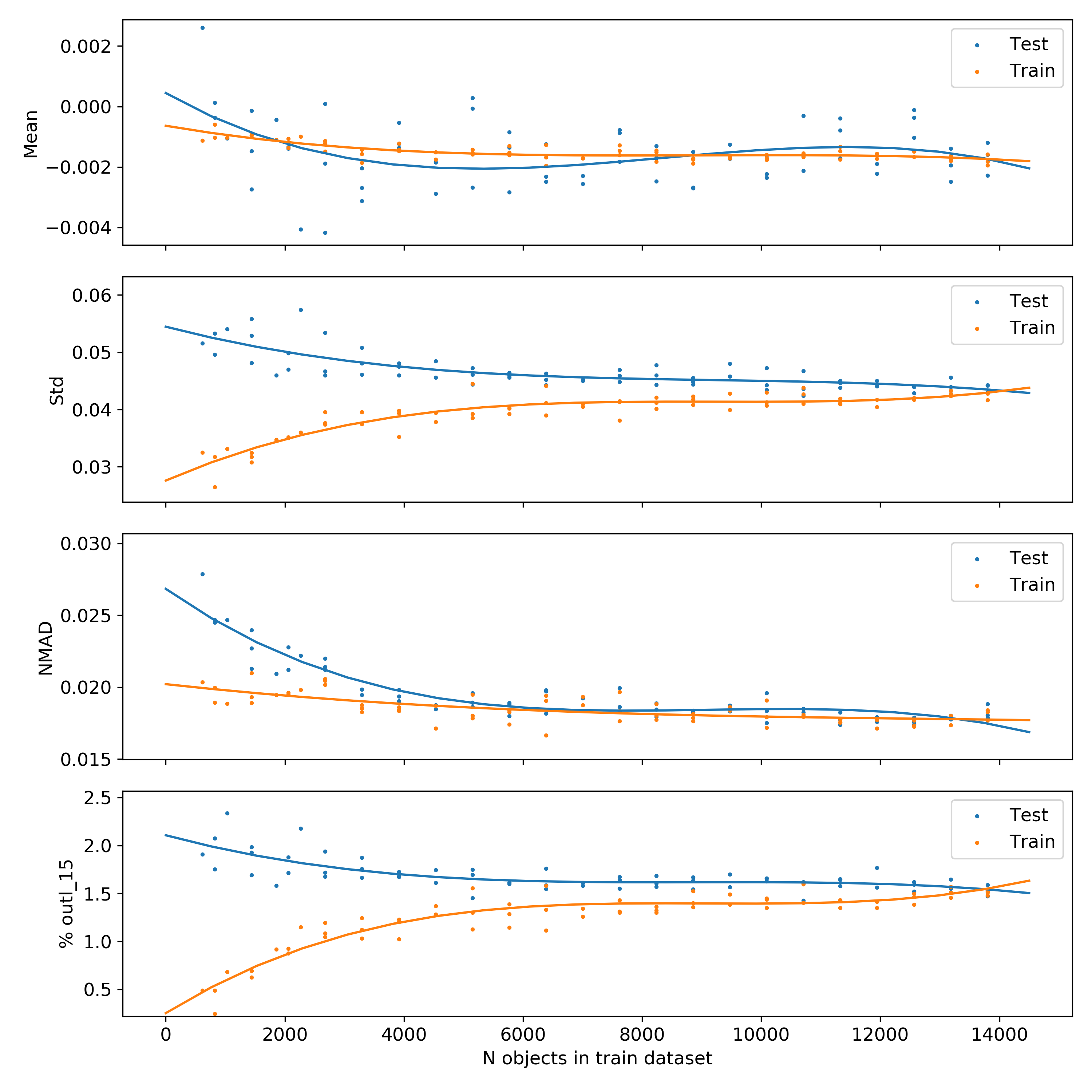}
     \caption{The change of quality of photo-z predictions with the increase of the size of the training sample. Using NMAD as the most stable metric we can see that the training sample as small as $\sim 6\,000$ objects provides enough generalization for the model.}
     \label{fig:learningCurve}
    \end{figure*}
    
    \begin{figure*}
    \centering
     \includegraphics[width=0.9\textwidth]{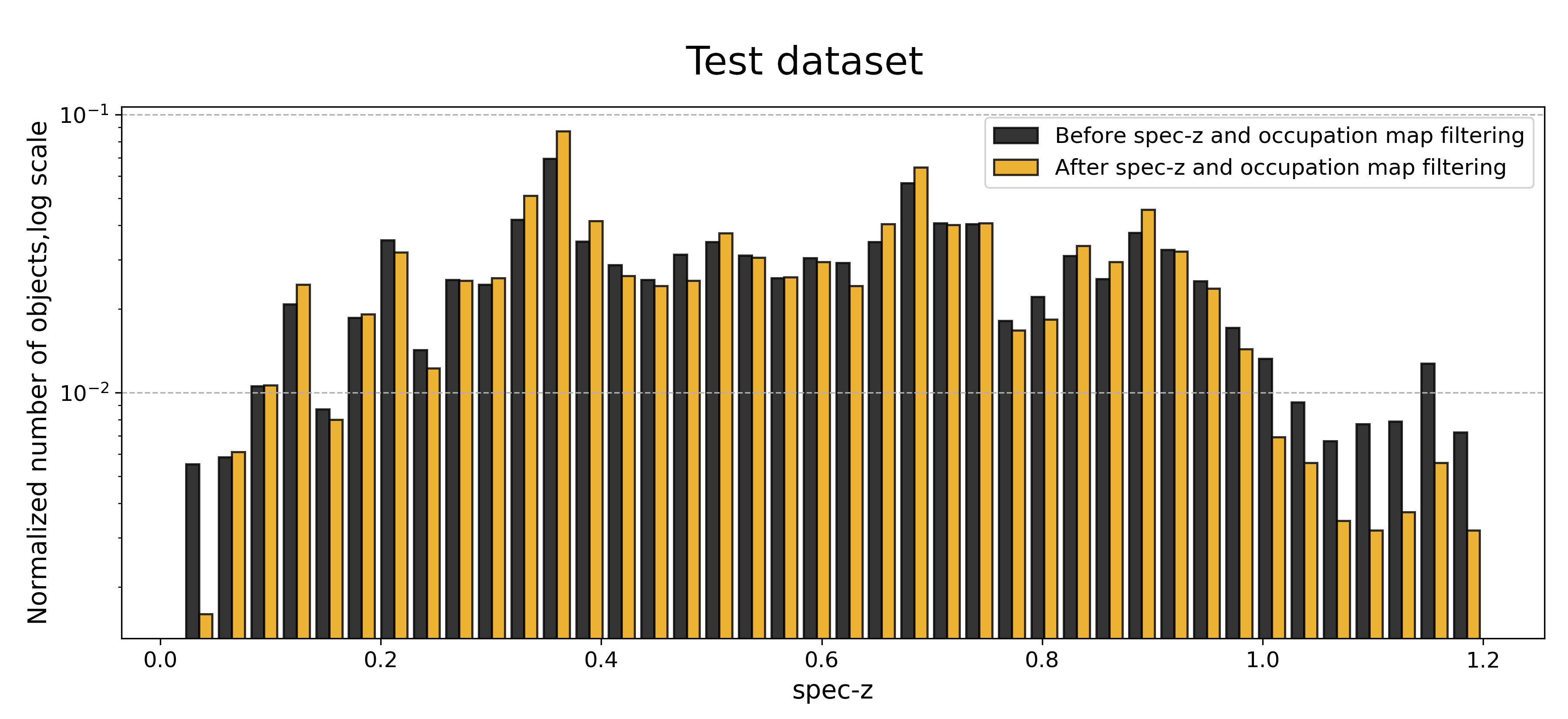}
     \includegraphics[width=0.9\textwidth]{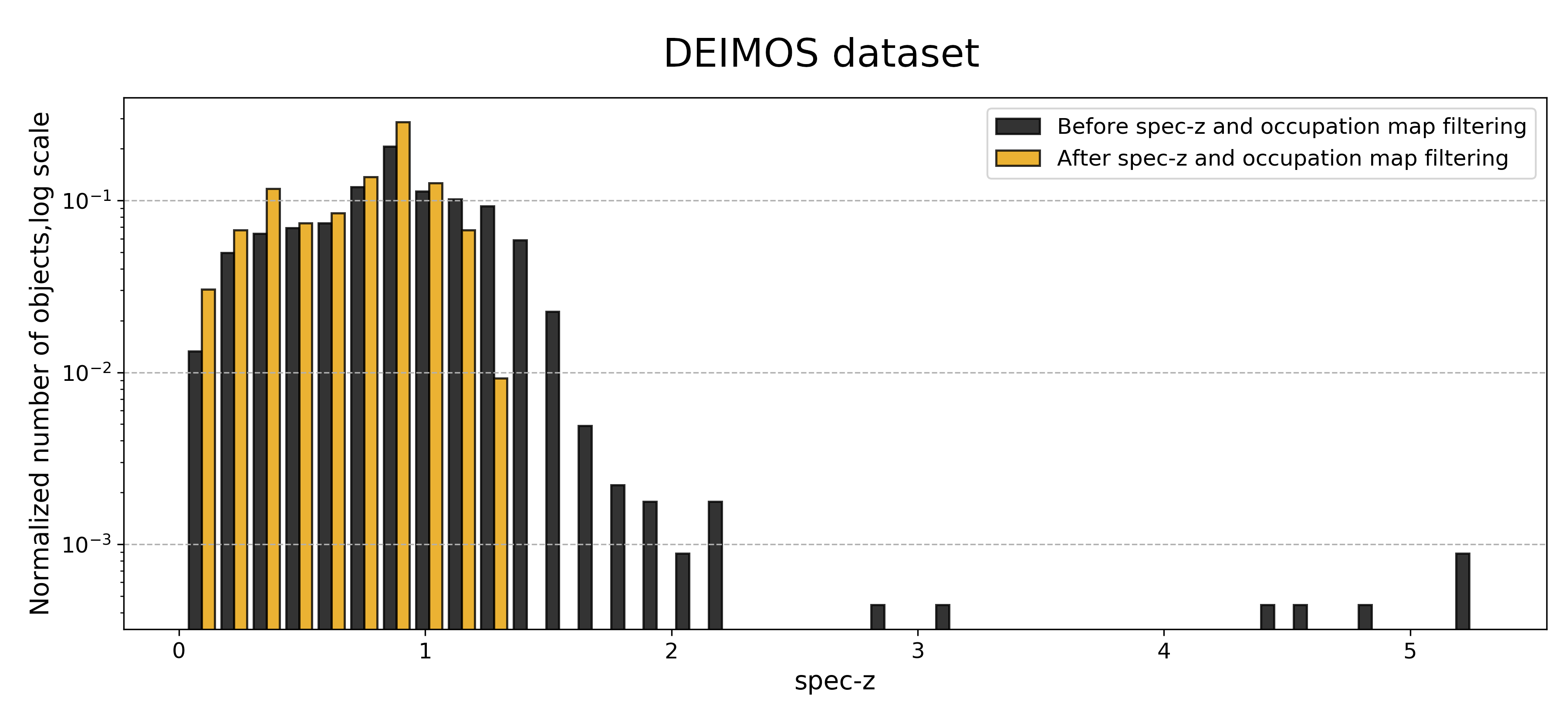}
     \caption{Spec-z distributions before and after SOM cleaning (using both $K_{spec}$ filtering and occupation maps cleaning). The normalized distribution of spec-z of the test sample (upper panel) does not change much.
     After the cleaning, there is a diminishing of the number of objects with $z_{spec} < 0.02$ and much softer decline of the relative share of objects with $1 < z_{spec} < 1.2$, but the shape of the middle part of the distribution remains intact.  The lower panel  shows that in the case of the DEIMOS, the SOM cleaning procedures do exactly what they are supposed to do: remove objects with $z_{spec}$ higher than 1.2, which was the limit for the KB.}
     \label{fig:specZDistribs}
    \end{figure*}
    
    \begin{figure*}
    \centering
     \includegraphics[width=0.82\textwidth]{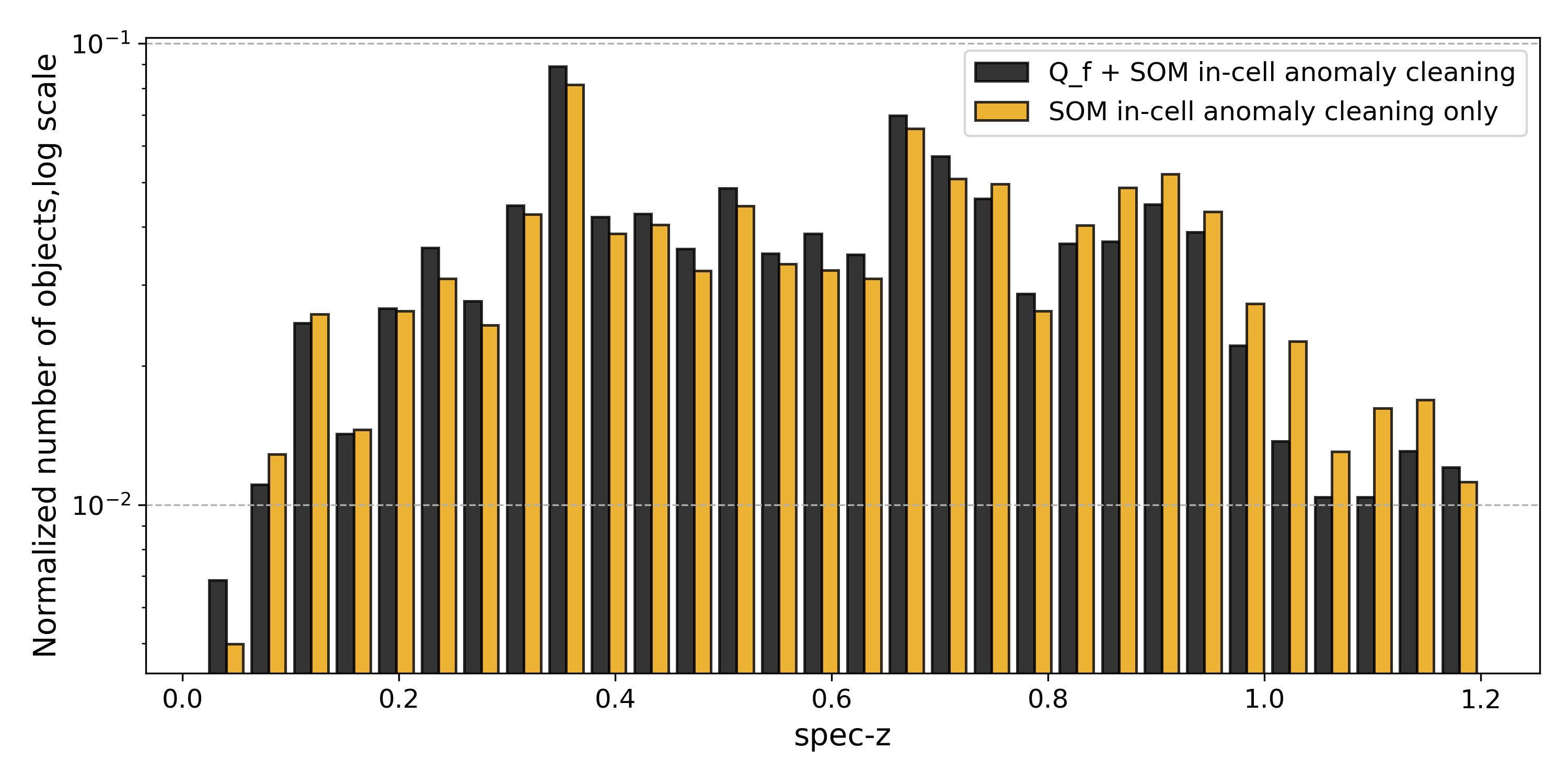}
     \caption{Spec-z distribution for the test dataset with and without standard \texttt{Q\_f} cleaning. The shapes of both distributions are very close, although SOM-only cleaning preserves slightly more high-redshift objects.}
     \label{fig:specZDistribDirty}
    \end{figure*}
    




\bsp	
\label{lastpage}
\end{document}